\documentclass[useAMS, usenatbib]{mnras}
\usepackage{times}
\usepackage{graphicx}
\usepackage{comment}
\usepackage{color}
\usepackage{amsmath}
\usepackage{amssymb}
\usepackage{datetime}
\usepackage{subfig}

\def\mnras{MNRAS}
\def\apj{ApJ}
\def\apjs{ApJS}
\def\nat{Nature}

\def \msun{\rm M_{\odot}}

\begin{document}

\title[Subhalo Demographics in Illustris]{Subhalo Demographics in the Illustris Simulation: \\Effects of Baryons and Halo-to-Halo Variation}

\author[K. T. E. Chua et al.]{Kun Ting Eddie Chua$^1$\thanks{Email: kchua@cfa.harvard.edu}, Annalisa Pillepich$^1$$^,$$^2$ , Vicente Rodriguez-Gomez$^1$$^,$$^3$, 
\newauthor
Mark Vogelsberger$^4$, Simeon Bird$^3$, and Lars Hernquist$^1$\\
$^{1}$Harvard-Smithsonian Center for Astrophysics, 60 Garden Street, Cambridge, MA 02138\\
$^{2}$Max-Planck-Institut f{\"u}r Astronomie, K{\"o}nigstuhl 17, 69117 Heidelberg, Germany\\
$^{3}$Department of Physics and Astronomy, Johns Hopkins University, 3400 N. Charles St., Baltimore, MD 21218, USA\\
$^{4}$Massachusetts Institute of Technology, Cambridge, MA 02138}

\maketitle


\begin{abstract}
We study the abundance of subhaloes in the hydrodynamical cosmological simulation Illustris, which includes both baryons and dark matter in a $\Lambda$CDM volume 106.5 Mpc a side.
We compare Illustris to its dark matter-only (DMO) analogue, Illustris-Dark, and quantify the effects of baryonic processes on the demographics of subhaloes in the host mass range $10^{11}$ to  $3 \times 10^{14} \msun$. We focus on both the evolved ($z=0$) subhalo cumulative mass functions (SHMF) and the statistics of subhaloes ever accreted, i.e. infall subhalo mass function. We quantify the variance in subhalo abundance at fixed host mass and investigate the physical reasons responsible for such scatter.
We find that in Illustris, baryonic physics impacts both the infall and $z=0$ subhalo abundance by tilting the DMO function and suppressing the abundance of low-mass subhaloes. The breaking of self-similarity in the subhalo abundance at $z=0$ is enhanced by the inclusion of baryonic physics. 
The non-monotonic alteration of the evolved subhalo abundances can be explained by the modification of the concentration--mass relation of Illustris hosts compared to Illustris-Dark.
Interestingly, the baryonic implementation in Illustris does not lead to an increase in the halo-to-halo variation compared to Illustris-Dark. In both cases, the normalized intrinsic scatter today is larger for Milky Way-like haloes than for cluster-sized objects. For Milky Way-like haloes, it increases from about eight per cent at infall to about 25 per cent at the current epoch. In both runs, haloes of fixed mass formed later host more subhaloes than early formers.
\\
\end{abstract}


\section{Introduction}

Observational data points to  a $\Lambda$CDM cosmology as the standard model of cosmic structure formation \cite[e.g.][and references therein]{Spergel03v148}.
Within this paradigm, primordial Gaussian fluctuations lead to the growth  of cold dark matter (DM) haloes, which form successively larger structures by accreting diffuse matter and merging with other haloes.
Although the accreted haloes are subject to various disruptive  processes (tidal stripping, tidal shocking and ram-pressure stripping etc.), many of these haloes are not completely destroyed,
surviving as gravitationally bound subhaloes in their host haloes.
By being able to resolve these substructures, early $N$-body cosmological simulations have been important in understanding the properties and evolution of subhaloes under different environments \citep[e.g.][]{Springel01v328,Gao04v355,Maccio06v366,Diemand07v667,Aquarius,Angulo09v399}.

Under the two-stage hierarchical formation scenario proposed by \cite{White78v183}, cooling of gas and subsequent star formation give rise to luminous galaxies that trace the underlying DM distribution.
Because $N$-body simulations do not track the evolution of baryons, different techniques have been developed to link the properties of these simulated DM haloes to observables such as the galaxy luminosity function.
These methods include semi-analytical modeling \citep[SAM, e.g.][]{Somerville99v310,Benson03v599,Bower06v370} and  abundance matching \citep[e.g.][]{Conroy09v696,Guo10v404}.
On the other hand,  $N$-body simulations are unable to provide a complete picture of galaxy formation, since baryons and DM are coupled and their co-evolution can have a significant impact on the structure of both DM haloes and subhaloes.

In early analytical work on the baryonic back-reaction on DM haloes, \cite{Blumenthal86v301} found that the condensation of baryons at halo centers can result in DM haloes that are more centrally concentrated, in a process known such as adiabatic contraction.
Indeed, radiative hydrodynamic simulations with baryonic cooling have been found to produce haloes that are both more spherical and centrally concentrated \citep[e.g.][]{Gnedin04v616,Abadi10v407}.
Another important factor observed in analytical models and simulations that play a role in altering the properties of DM haloes is baryonic feedback.
In contrast to adiabatic contraction, simulations including stellar and AGN feedback have observed that the back-reaction from baryons can lead to lower DM central densities and halo concentrations compared to DM-only (DMO) simulations \citep[e.g.][]{Duffy10v405}.
Because these baryonic processes are difficult to model analytically, numerical simulations are key in understanding how dark matter structure can be affected by baryons.

While subhaloes have been heavily studied using $N$-body simulations, there have been few hydrodynamic studies conducted, due to  the high computational power and memory needed to compute additional baryonic processes.
Initial hydrodynamic studies of subhalo abundances focused on a small range of halo mass, where individual haloes were simulated via resimulation techniques.
For example, \cite{Dolag09v399} performed zoom-in hydrodynamic resimulations of cluster-size haloes and found that radiative processes are important in determining the subhalo abundance.
In non-radiative simulations, they found that gas in subhaloes is easily stripped by ram pressure, suppressing subhalo abundance compared to $N$-body simulations.
On the other hand, in radiative simulations that include feedback, they found that star formation can lead to subhalo abundances that are similar to or even exceed those of the $N$-body counterparts.

For Milky Way (MW)-mass haloes, the zoom-in resimulations of \cite{Wadepuhl11v410} observed that the effect of hydrodynamic processes varies non-monotonically with subhalo mass.
They showed that baryonic processes can suppress the abundance of small subhaloes, while simultaneously increasing the abundance of massive subhaloes compared to DM-only runs.
Similar results have also been found with {\sc arepo}-simulated Milky Way zooms \citep{Zhu16v458} and the Eris Simulation \citep[][Annalisa Pillepich, private communication]{Guedes11v742}
Previously, \cite{Weinberg08v678} had also noted the enhanced survival of subhaloes in hydrodynamical simulations, although they did not observe any suppression in halo occupation, most likely due to their low-mass resolution ($m_{\rm DM} = 7.9 \times 10^8\, M_\odot$).
\cite{Maccio06v366} had also found from a resimulation of a MW-mass halo that the number density of substructure is enhanced up to an order of magnitude in the inner regions of the halo.

The varied results from these zoom-in simulations show that the effect of baryons on subhalo abundance is complex, and there has been no consensus as to whether the abundance of subhaloes is suppressed or increased, and how this can depend on the properties of the host haloes or the subhaloes.

The diversity of the effect of baryonic processes is not limited to subhalo abundances. 
In \cite{Vogelsberger14v444}, it was found that the relative subhalo masses between the hydrodynamic cosmological simulation Illustris and its DMO counterpart Illustris-Dark exhibits a peak at $10^{11} M_\odot$. This result is not shared by other hydrodynamic simulations.
For example, \cite{Sawala13v431} found from the hydrodynamical simulation GIMIC \citep{Crain09v399} that baryons reduce the mass of subhaloes less massive than $10^{12} M_\odot$, and the relative subhalo masses varies monotonically between  0.65 and 1 for subhaloes of mass $10^{9} \msun$ and  $10^{14} \msun$ respectively. Additionally, the subhalo abundance is suppressed in GIMIC at small halo masses but never exceeds the $N$-body run for large halo masses. A similar result has also been observed by \cite{Despali2016} in the EAGLE simulation \citep{EAGLE}.

While the previous analyses largely focused on the current subhalo abundance, the total number of subhaloes accreted by a halo over its lifetime is also an important quantity. Both \cite{Giocoli08v386} and \cite{Jiang16v458} found in their analytic models that this {\it infall} subhalo mass function (SHMF) is independent of halo mass, in contrast to the current (or {\it evolved}) SHMF.
The DMO infall SHMF is a `pure' quantity in the sense that a subhalo before accretion is not subject to stripping and can be calculated directly from an $N$-body merger tree algorithm.
However, the analytic model of \cite{Jiang16v458} excludes baryonic physics. Clearly, we feel that it is important to also understand the effect of baryons on the infall subhalo abundance using hydrodynamical simulations. To our knowledge, such a study has not been performed previously.

In this work, we aim to investigate and quantify the effect of baryon physics on the properties of subhaloes by comparing the hydrodynamical simulation Illustris with its DM-only counterpart Illustris-Dark.
With a large box size (106.5 Mpc a side) and high resolution, Illustris simulates a large number of haloes ranging from $10^{11} M_\odot$ to a few $10^{14} M_\odot$,
thus enabling a statistical analysis of subhalo properties.
With these simulations, we are able to observe the impact of baryons and feedback across a wide range of subhalo and halo masses. 
More importantly, we will characterize in addition to the average final subhalo abundance, the subhalo abundance at infall and the halo-to-halo variation of subhalo abundances.

This paper is organized as follows:
Section 2 provides a description of the simulations and methods that were used for this work.
We describe the impact of baryons on the average cumulative subhalo mass function in Section 3, 
and the halo-to-halo variation of subhalo abundances in Section 4.
We explain the reasons behind baryonic effects on the average subhalo abundance in Section 5, and discuss the physical origins of the baryonic effects and the scatter in Section 6.
Finally, we summarize our conclusions in Section 7.

\section{Methods and Definitions}
\subsection{The Illustris Project Simulations}

The analysis presented here is based on the Illustris project \citep{Illustris,Vogelsberger14v444,Genel14v445, Sijacki15v452}, a series of cosmological simulations encompassing a volume 106.5 Mpc a side and evolved in a $\Lambda$CDM cosmology consistent with WMAP-9 results \citep[][namely $\Omega_m = 0.27$, $\Omega_\Lambda = 0.73$, $\Omega_b = 0.0456$, $\sigma_8 = 0.81$, $n_s = 0.963$, $h = 0.704$]{Hinshaw2013}. 

The suite includes three realizations at different resolutions including gravity, hydrodynamics and key physical processes for galaxy formation. The highest resolution run -- Illustris-1, hereafter Illustris or full-physics (FP) run --, follows  $1820^3$ DM particles and $1820^3$ gas cells, with a mass resolution of $6.26 \times 10^6 \msun$ and $1.26 \times 10^6  \msun$ (for DM and baryons, respectively). The comoving gravitational softening lengths at $z=0$ are 1.4 and 0.7 kpc for DM and baryonic collisionless particles, respectively. The gas gravitational softening length is adaptive and set by the cell size, with a floor given by the aforementioned 0.7 kpc, however, the sizes of the cells used to evolve the gas can be much smaller than this. The two lower resolution simulations (Illustris-2 and 3) have mass resolutions 8 and 64 times lower, and softening lengths 2 and 4 times larger.  
For comparison to the full-physics runs, an analog series of DM-only (DMO) simulations are available, with the same initial conditions and corresponding resolution. The highest resolution dissipationless  DMO run is called Illustris-1-Dark (hereafter Illustris-Dark), and evolves $1820^3$ DM particles and no baryonic resolution elements.

All simulations are carried out using \textsc{arepo} \citep{Springel09v401}, where a moving-mesh is used to solve the hydrodynamical equations with a finite volume approach.
The resulting method is quasi-Lagrangian and combines the advantages of previous Eulerian and Lagrangian schemes, giving rise to a scheme that is highly reliable and adaptive \citep{Vogelsberger12v425,Sijacki12v424}. The gravitational forces are calculated with a Tree-Particle-Mesh (Tree-PM) scheme, which maintains all the most important advantages of the tree algorithm: its insensitivity to clustering, its essentially unlimited dynamic range, and its precise control of the softening scale of the gravitational force. Importantly, the combination of the particle-mesh method with the tree algorithm delivers much higher accuracy than the pure particle-mesh methods usually adopted within adaptive mesh refinement (AMR) codes. This makes the Tree-PM a superior choice particularly in the case of cosmological simulations, where mesh-based Poisson solvers typically implemented in AMR codes have been shown to underestimate structure formation at the low-mass end of the dark-matter halo mass function \citep{OShea05v160, Heitmann08}.

The galaxy formation implementation is fully described in \cite{Vogelsberger13, Torrey14v438} but we review the relevant features here. Since the large-scale structure simulation lacks the spatial and mass resolution to resolve the small-scale baryon physics, relevant baryonic processes have to be treated using sub-resolution models that link the scales that are actually resolved by the simulation to the unresolved processes. As such, star formation is modelled following \cite{Springel03v339} where the star forming interstellar medium is described using an effective equation of state and stars form stochastically above a gas density
$\rho_{\rm sfr} = 0.13\,{\rm cm^{-3}}$ with timescale $t_{\rm sfr} = 2.2{\rm\,Gyr}$. To avoid the over-cooling that used to plague early numerical simulations, galactic winds and feedback from active galactic nuclei (AGN) are adopted to quench star formation in Illustris. The stellar feedback is modeled as kinetic outflows. The AGN feedback includes both quasar-mode and radio-mode energy releases into the surrounding gas according to the central black hole accretion rate,  as well as non-thermal and non-mechanical electromagnetic feedback. The subgrid parameters have been chosen to reproduce the observed cosmic star-formation rate density, the current galaxy stellar mass function, and the stellar mass - halo mass relation of galaxies at $z=0$. Indeed, Illustris has demonstrated excellent to reasonable agreement with a broad number of observational scaling relations and galaxy properties at low redshift  \citep{Vogelsberger14v444} as well as across cosmic time \citep{Genel14v445}.

\subsection{(Sub)Halo Identification and Mass Definitions}
\label{subsec:defs}

Haloes, subhaloes, and their basic properties are obtained with the {\sc FOF} and {\sc subfind} algorithms \citep{Davis85, Springel01v328,Dolag09v399}, at each of the 136 stored snapshots from $z\sim 40$ to $z=0$. First a standard friends-of-friends group finder is executed to identify {\sc FOF} haloes (linking length 0.2) within which gravitationally bound substructures are then located and characterized hierarchically. The {\sc subfind} catalog therefore includes both central and satellite subhaloes: the former are independent particle associations which may contain other subhaloes and whose center coincides with the {\sc FOF} center; the latter may be either dark or luminous, and are members of their parent {\sc FOF} group regardless of their distance from the {\sc FOF} centre. Also imposed are particle number cuts (20 and 32 total resolution elements per (sub)halo for the {\sc FOF} and {\sc subfind} catalogs respectively). For the halo masses examined in this work ($M_{200}> 10^{11} \msun$), there is a one-to-one correspondence between {\sc FOF} haloes and {\sc subfind} haloes which are centrals. 

Throughout this paper, we refer to subhaloes that do not reside within $R_{200}$\footnote{$R_{\Delta}$ is defined as the radius within which the mean enclosed mass density is $\Delta$ times the critical value $\rho_c$ i.e. $\bar \rho_{\rm halo} = \Delta \rho_c$.} of a larger halo as {\it host, central} or {\it parent}  haloes. 
Unless otherwise noted, any central halo is associated to only those subhaloes identified within its virial radius, $R_{200}$. In particular, we note that this follows the definition utilized by \cite{Gao11v410}, but differs from \cite{BK10v406,Bosch16v458}.

The evolution of hosts and subhaloes is followed across cosmic times with the {\sc sublink} merger tree \citep{Vicente2015v449}. There, the main branch of any (sub)halo at $z=0$ is defined as the sequence of progenitors with the most massive history behind them (rather than the sequence of progenitors which maximize the mass at every time step). The time a subhalo first enters the virial radius of its $z=0$ host is called the {\it infall} or accretion time. 

Throughout, host haloes are characterized by the spherical-overdensity mass $M_{200}$, obtained by summing the mass of {\it all} particles and resolution elements enclosed within $R_{200}$ (so including DM, gas elements, stars and black holes).
At $z=0$, we characterize subhaloes by their total mass, $m_{\rm sub}$, obtained by summing the mass of all gravitationally bound particles to the subhalo according to {\sc subfind}, regardless of their distance. Analogously, the mass of a subhalo at infall ($m_{\rm acc}$) is the sum of the mass of all resolution elements bound to the subhalo at its infall time. 

Note that these mass definitions imply different total masses for hosts and subhaloes which contain the same number of DM particles in Illustris and Illustris-Dark. Yet, this choice shall be preferred to e.g. the labeling of (sub)haloes based solely on their DM mass, as this cannot be related to observations and the physical mechanisms dominating the host-subhalo interactions depend on the total mass. Moreover, previous works have demonstrated that the peak height of the circular velocity curve ($V_{\rm max}$) is a good descriptor of subhaloes as it is less prone to variations than mass across their life time and orbital phases around hosts \citep{Diemand07v667, Kuhlen07v671}. The abundances of subhaloes have even been quantified based on the $V_{\rm peak}$, the largest value of $V_{\rm max}$ a subhalo has ever had along its evolutionary path (and usually met before a subhalo falls into the potential well of a more massive one, because of the effects of tidal stripping). However, these choices applies only to DMO simulations: when baryonic physics is included, the physical meaning of $V_{\rm max}$ can change dramatically, as the peak height of the circular velocity of a (sub)halo can be dominated by the stellar component of the object (e.g. the bulge of a galaxy) rather than the DM one and can change because of star formation rather than tidal stripping: this makes the comparison between e.g. $V_{\rm max}$ in Illustris and $V_{\rm max}$ in Illustris-Dark difficult to interpret, and the total mass a better choice  \citep[however, see][for the effects of baryons on the subhalo velocity function]{Zhu16v458}.

\subsection{Matching Subhaloes}
\label{subsec:matching}

To quantify comparisons between Illustris and Illustris-Dark and isolate the effects of baryons, we match host haloes between the two simulations in order to identify DMO and full-physics (FP) {\it analogs}: we do so by using the unique IDs of their DM particles. The precise strategy is described in details in \cite{RodriguezGomez2016} and is based on the {\sc subfind} catalog only. In practice, for any given (sub)halo in Illustris, the matched (sub)halo in Illustris-Dark is the (sub)halo which contains the largest fraction of common IDs among the most bound  DM particles. The same can be done starting from a (sub)halo in Illustris-Dark to find a match in Illustris.
The final matched catalogues consist of those objects which have a successful bijective match between Illustris and Illustris-Dark. We use the term {\it disrupted} to refer to (sub)haloes that are either missing in Illustris compared to Illustris-Dark (or vice versa) or fall below our imposed subhalo resolution limit. For massive central haloes it is very rare to encounter unmatched cases: out of 14311 haloes in Illustris with $M_{200} > 10^{11}\msun$, 14306 of them are bijectively matched to Illustris-Dark objects.

Finally, we match (sub)haloes between Illustris and Illustris-Dark at all fixed snapshots, but without imposing consistency between Illustris and Illustris-Dark merger tree branches. It might happen, for example, that the DMO analogs of the main branch elements of an Illustris host at $z=0$ do not lie along the main branch of the analog host in Illustris-Dark at $z=0$. Moreover, evolutionary tracks of (sub)haloes in Illustris and Illustris-Dark may differ or may be shifted in cosmic time: so it can happen that the infall time of a host-subhalo pair in Illustris does not coincide with the infall time of the matched pair in Illustris-Dark, adding a contribution to the differences between masses of hosts and subhaloes in the full-physics and DMO runs. These effects will all contribute to the overall discrepancies between Illustris and Illustris-Dark but we will not attempt to isolate them.

\subsection{Subhalo Mass Function (SHMF): $z=0$ and Infall}

The quantity we use in this paper to describe the abundance of subhaloes is simply the cumulative subhalo mass function (SHMF). At a given time, this is the number of subhaloes that a parent halo of a given mass hosts above a given subhalo {\it mass fraction} $\mu$.

At $z=0$, the mass fraction reads $\mu \equiv m_{\rm sub}/M_{200}$, and the abundance $N(> \mu)$ can be measured for any given host halo in the simulation. By grouping the hosts in mass bins, the SHMF is usually reported as the average (the mean) subhalo number across hosts. We refer to this quantity as the {\it current or evolved} SHMF, measured using the subhaloes in the final snapshot of the simulation that have survived within the host gravitational field to the present epoch after being accreted and that pass the resolution limit threshold.

Following \cite{Bosch05v359} and \cite{Jiang16v458} (hereafter JB16), we also measure the {\it infall or unevolved} SHMF: this includes all subhaloes that have ever been accreted onto a host since its formation. The infall SHMF therefore accounts for subhaloes that may get totally disrupted (`killed') after infall, subhaloes on elongated orbits that put them outside $R_{200}$ at $z=0$, and also subhaloes that have been stripped below the adopted mass resolution limit. 
Here we focus on the infall SHMF of all hosts identified at $z=0$, for each of which a unique main branch (or evolution history) is given by a merger tree (see Section \ref{subsec:defs}). Throughout, for each unique host across cosmic time, we identify all subhaloes that have ever entered within the host's evolving virial radius $R_{200}$; we then normalize the cumulative function with the subhalo mass fraction $\mu_{\rm acc} = m_{\rm acc}/M_{200}$, where $m_{\rm acc}$ is the total mass of the subhalo at the infall time and $M_{200}$ is the total mass of the host {\it at $z=0$}. Also the infall SHMF is given as the cumulative number of subhaloes averaged across groups of host haloes, e.g. bins in host mass.

In what follows, beyond the average, we will characterize the subhalo abundances also by means of other summary statistics for the distribution of $N(>\mu)$ or $N(>\mu_{\rm acc})$ for a sample of host haloes, e.g. the standard deviation and the lower and upper quartiles. 

Finally, the infall SHMF curves would be slightly different than what is shown in the next sections if we chose to define infall as the time at which the total subhalo mass is maximized (usually before crossing the virial radius). However, we believe that none of the conclusions and relative statements between Illustris and Illustris-Dark would be different.

\subsection{Resolution Limits}
\label{subsec:limits}

The minimum subhalo mass we report our results for has been determined by comparing the evolved and unevolved SHMFs of Illustris-Dark to the lower resolution runs Illustris-Dark-2 and Illustris-Dark-3. In practice, we find that a minimum of 160 DM particles in Illustris-Dark is required to avoid incompleteness in both the current and infall subhalo abundance: in Illustris-Dark, this corresponds to a minimum subhalo mass of $10^9 M_\odot$.

In what follows, we impose everywhere the same minimum subhalo total mass of $10^9 M_\odot$ in both Illustris and Illustris-Dark (and scale the limit upwards for the lower resolution runs). For Milky Way-like hosts, this implies that Illustris can reliably provide statistics for subhaloes with normalized mass of about $\mu \gtrsim 10^{-3}$. It should be noted that the same minimum mass in fact corresponds to (sub)haloes resolved with a larger number of resolution elements in the FP run with respect to the DMO cases, and hence our minimum mass choice is somewhat conservative for Illustris. Moreover, the same minimum mass at both $z=0$ and infall time also implies some differences and subtleties between the two cases, as the same mass cut means that a subhalo population at $z=0$ may not include all the descendants of the subhalo population at infall. 

By keeping in mind these complications, we think that a universal mass cut is the easiest choice to justify and discuss the results we present in the following sections.\\

\subsection{Other Host Halo Properties}
\label{subsec:hostprops}

In addition to the host mass, in our analysis we will make use of other properties to characterize the host haloes: particularly, the halo formation redshift, concentration and DM halo shapes. 

\begin{itemize}
\item {\it Halo Formation Redshift}: it denotes the redshift when a halo has accreted half of its $z=0$ total mass. We measure it from the simulated haloes similar to \cite{Bray16v455}, as the earliest moment in cosmic time at which the splined total mass accretion history of a halo reaches 50 per cent of its current-epoch mass. \\

\item {\it Halo Concentration}: it characterizes the DM density profile of a halo. Halo concentrations are obtained by fitting the DM density profile of each halo $\rho_{\rm DM}(r)$ to an Einasto profile \citep{Einasto,Navarro04v349}:
\begin{equation} 
\rho_{\rm DM}(r) = \rho_{-2} \exp\left\{ -2n \left[ \left( \frac{r}{r_{-2}} \right)^{1/n} -1 \right]\right\},
\end{equation}
where $n$ indicates the sharpness of the profile, while $\rho_{-2}$ and $r_{-2}$ indicates the radius and density where the slope of the density profile takes on the isothermal value i.e. $d \ln \rho/d\ln r =-2$. The concentration parameter is then defined as $c_{-2}\equiv R_{200}/r_{-2}$. This definition is slightly different from the conventional one based on the Navarro-Frenk-White profile functional form \citep{Navarro96v462}, but it is a more general choice for cases when DM haloes deviates from NFW profiles, as may be the case in hydrodynamical simulations \citep[e.g.][]{Pedrosa09v395}. \\

\item {\it DM Halo Shape}: it summarizes the ellipsoidal shape of DM haloes in terms of the axis ratio $c/a \equiv s$ of the DM density ellipsoid (see Chua et al. in prep). In practice, we use an iterative procedure to determine $s$ in elliptical shells around the halo center using the shape tensor:
\begin{equation}
S_{ij} = \frac{\sum_k  m_k\, r_{k,i} \,r_{k,j}}{\sum_k m_k}
\end{equation}
where $m_k$ is the mass of the $k$th DM particle and $r_{k,i}$ is the $i$th component of its position. In each iteration, the shape tensor is first diagonalized to compute the eigenvectors and eigenvalues. The eigenvectors denote the directions of the principal axes and are used to rotate the positions of the particles into the principal frame. The eigenvalues  $a$, $b$ and $c$ ($a \ge b \ge c$) are used to compute the axis ratios $q \equiv b/a$ and $s$.
Starting from $q=1$ and $s=1$, the iteration is repeated until $q$ and $s$ converge i.e. when the values in successive iterations change by less than one per cent.
For parent haloes, we exclude the contribution from DM particles which are gravitationally bound to its subhaloes \citep[see e.g.][]{Zemp11v197}. In what follows, we quote the shape parameter in the volume around $0.15\times R_{200}$. This choice is motivated by previous numerical simulations that showed baryonic effects to be stronger in the inner halo where galaxies reside, leading to substantially rounder inner haloes in hydrodynamic simulations compared to their DMO counterparts \citep[e.g.][]{Abadi10v407,Butsky2016v462}.

\end{itemize}

\section{Average Subhalo Abundances}

\subsection{Cumulative Subhalo Mass Functions}
\label{sec:shmf}

\begin{figure*}
  \centering
  \includegraphics[width=\textwidth]{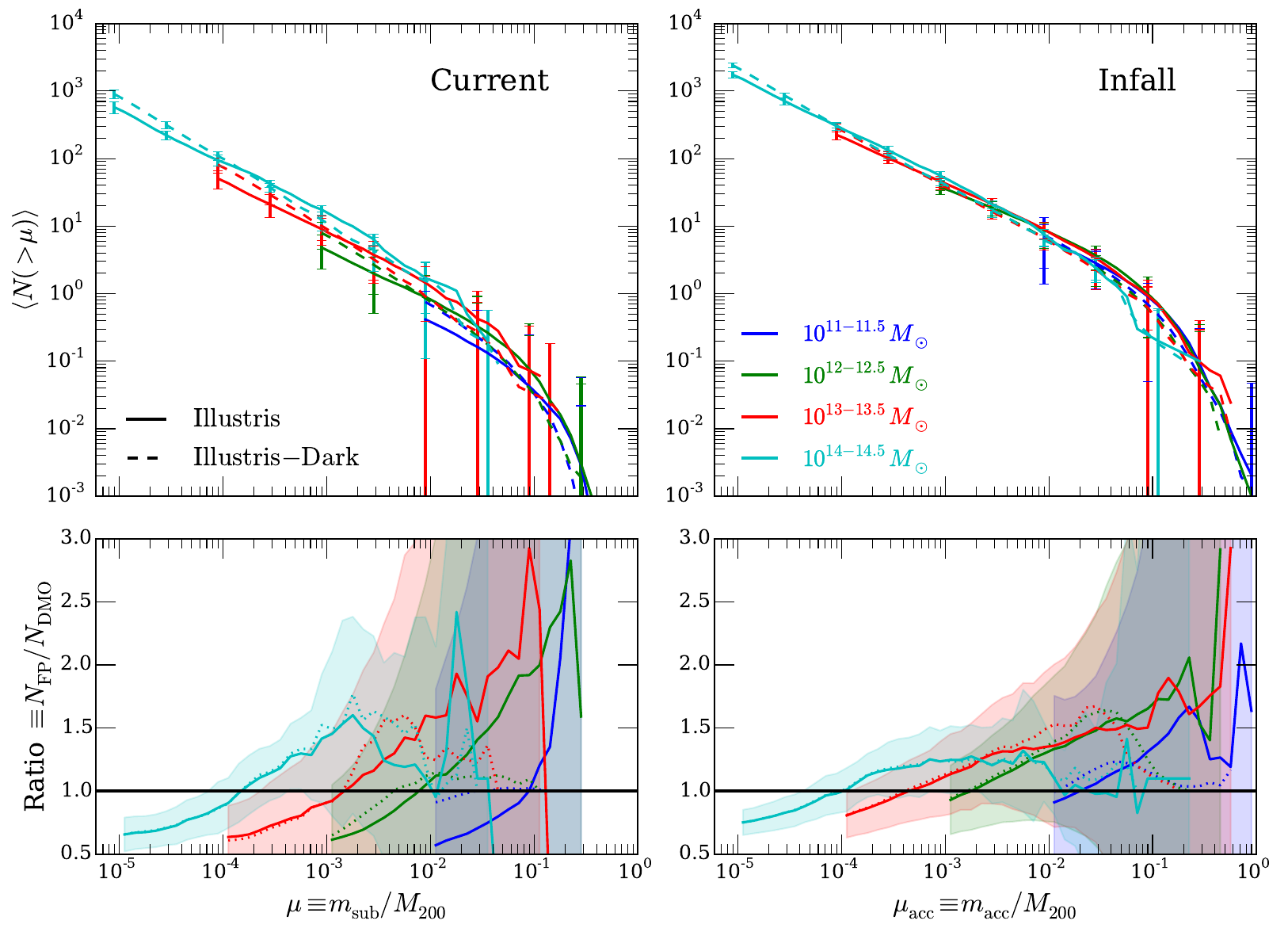}
  \caption{Subhalo abundances in two cosmological simulations.
    Upper panels: Illustris and Illustris-Dark cumulative subhalo mass functions (SHMF) plotted as a function of the normalized subhalo mass ($\mu$ and $\mu_{\rm acc}$) at present time (left) and at infall (right). 
    Solid and dashed lines represent results from Illustris and Illustris-Dark respectively.
    Coloured lines correspond to different host halo masses. 
    The error bars show the standard deviation in each bin for Illustris and are spaced to reduce correlation.
    Best fit parameters of each SHMF to equation \ref{eq:cdf} are shown in Table \ref{table:fit}.
    Lower panels: Effects of baryonic physics. Solid curves: mean ratio of the SHMF in Illustris to that of Illustris-Dark  as a function of the normalized subhalo mass. Dotted curves denote the same ratio for the subhalo abundances in Illustris to that of only {\it matched host haloes} in Illustris-Dark. The shaded regions are the propagated uncertainties using the standard deviation around the mean ratio.
  }
\label{fig:shmf}
\end{figure*}

Results for the current and infall SHMFs are given in Figure \ref{fig:shmf}, upper panels: solid and dashed curves denote the average subhalo mass function from Illustris and Illustris-Dark, respectively. 
Mean subhalo numbers are given in bins of host masses, as indicated by the colours of the curves: namely, we have considered halo masses of $10^{11-11.5} M_\odot$, $10^{12-12.5} M_\odot$,
$10^{13-13.5}  M_\odot$  and $10^{14-14.5} M_\odot$. The number of host haloes identified in Illustris (Illustris-Dark) within these mass intervals are 
9690 (8720), 992 (1052), 82 (120) and 10 (11) respectively.
The error bars (plotted only for Illustris) show the standard deviation around the mean number of haloes per host bin, plotted at intervals to reduce correlation between neighbouring points.

As expected, at all host and subhalo masses, the subhalo abundances at infall are larger than at $z=0$, as a consequence of the host-subhalo interactions and the subsequent mass loss or even disruption of subhaloes after infall. However, we also observed a non negligible discrepancy between the subhalo abundances with and without the effects of baryonic physics, at all times. In all cases (Illustris and Illustris-Dark, at both $z=0$ and infall), the shape of the SHMFs is consistent with a power law for low-mass subhaloes which degrades into a steep drop-off for the abundance of relatively very massive subhaloes.
Following \cite{BK10v406}, this can be fit with the formula: 
\begin{equation}
 \left<N(>\mu)\right>_{\rm fit} = \left(\frac{\mu}{\tilde \mu_1}\right)^a \exp \left[ -\left( \frac{\mu}{\mu_{\rm cut}}\right)^b\right]
\label{eq:cdf}
\end{equation}
where $a$ represents the logarithmic slope at small $\mu$, $\tilde \mu_1$ controls the normalization, while $\mu_{\rm cut}$ and $b$ determine the position and steepness of the drop-off at large subhalo masses.

Best fit parameters for our SHMFs are given in Table \ref{table:fit}.
Since the fitting parameters are correlated with each other, we fix the value of $b$ to $b=1.3$ at current time and $b=0.8$ at infall, which produced reasonable fits for both Illustris and Illustris-Dark (see Appendix \ref{sec:shmffit} for more details). Our value of $b$ at current time is similar to that used in \cite{Gao11v410}, who fixed $b=1.2$.
To take into account the larger scatter at large $\mu$ and to better measure the logarithmic slope $a$, we minimized
\begin{equation}
\chi^2 = \sum_i \left( (N_i-N_{i,\rm fit})/\sigma_i \right)^2
\label{eq:chi2}
\end{equation}
where the sum runs over number of bins, $N_i$ is the mean of the SHMF, $N_{i,\rm fit}$ is the expression given in Equation \ref{eq:cdf} and $\sigma_i$ is the 1-$\sigma$ scatter (standard deviation).
We further calculate the uncertainty in the fit parameters using a bootstrap analysis.
The discrepancy between the subhalo abundances with and without the effects of baryonic physics can be also observed in the fitting parameters.

\begin{table*}
\begin{center}
\begin{tabular*}{0.9\textwidth}{@{\extracolsep{\fill}}c | c c c | c c c }
\multicolumn{1}{c}{} & \multicolumn{3}{c|}{\bf Illustris-Dark} & \multicolumn{3}{c}{\bf Illustris} \\ 
$M_{200}\,[M_\odot] $ & $\tilde{\mu_1}$ & $a$ & $\mu_{\rm cut}$&  $\tilde{\mu_1}$ & $a$ & $\mu_{\rm cut}$ \\
\hline
{\bf Current} & & & & & &\\ 
$10^{[11,11.5]}$ & $0.0070\pm0.0004$ & $-0.911\pm0.072$ & $0.103\pm0.017$ &
					$0.0031\pm0.0004$ & $-0.763\pm0.060$ & $0.147\pm0.027$ \\
$10^{[12,12.5]}$ & $0.0080\pm0.0003$ & $-0.972\pm0.016$ & $0.099\pm0.013$&
					$0.0078\pm0.0003$ & $-0.717\pm0.015$ & $0.107\pm0.010$ \\
$10^{[13,13.5]}$ & $0.0101\pm0.0006$ & $-0.945\pm0.013$ & $0.059\pm0.019$ &
					$0.0170\pm0.0002$ & $-0.748\pm0.015$ & $0.064\pm0.021$ \\
$10^{[14,14.5]}$ & $0.0136\pm0.0013$ & $-0.935\pm0.021$&  $0.051\pm0.034$ &
					$0.014\pm0.003$ & $-0.752\pm0.014$ & $0.016\pm0.008$  \\\
%
& & & & & & \\
{\bf Infall} & & & & & \\ 
$10^{[11,11.5]}$ & $0.025\pm0.003$ & $-0.705\pm0.058$ & $0.106\pm0.021$ &
				  $0.012\pm0.001$ & $-0.559\pm0.062$ & $0.103\pm0.020$ \\
$10^{[12,12.5]}$ & $0.039\pm0.002$ & $-0.781\pm0.011$ & $0.082\pm0.008$&
				 $0.048\pm0.003$ & $-0.607\pm0.011$ & $0.085\pm0.008$ \\
$10^{[13,13.5]}$ & $0.059\pm0.003$ & $-0.793\pm0.007$ & $0.057\pm0.008$ &
				  $0.101\pm0.009$ & $-0.669\pm0.010$ & $0.055\pm0.008$ \\
$10^{[14,14.5]}$ & $0.065\pm0.005$ & $-0.839\pm0.007$&  $0.063\pm0.015$ &
				 $0.202\pm0.018$ & $-0.701\pm0.006$ & $0.018\pm0.002$  \\
\hline
\end{tabular*}
\caption{Best fit parameters to equation \ref{eq:cdf} for the subhalo mass functions in Illustris and Illustris-Dark at current time (top) and infall (bottom).
The parameter $b$ has been fixed to $b=1.3$ at current time and $b=0.8$ at infall.
The uncertainties were obtained from a bootstrap analysis. In general, the fit parameters vary monotonically with host mass, with the widest mass trends appearing for Illustris at $z=0$, and the power-law slopes $a$ being shallower in Illustris than in Illustris-Dark (see text for a detailed discussion).}
\label{table:fit}
\end{center}
\end{table*} 

We note that our DMO results compare favourably to previous subhalo abundance measurements from simulations with similar volume and resolution. Let us consider for example results from the Millennium II simulation \citep[MS-II hereafter,][]{BK09v398}: \cite{Gao11v410} analyzed and provided fitting formulae for the SHMF in MS-II using the same subhalo finder adopted here across halo masses; \cite{BK10v406} focused on Milky Way mass hosts. In general, the subhalo abundance (normalization) is slightly larger in Illustris-Dark compared to MS-II ($\tilde{\mu_1} \sim 0.005-0.014$ vs. $\tilde{\mu_1} \sim 0.0085-0.011$), probably owing to the larger value of $\sigma_8$ in MS-II \citep[see ][for discussions on the effects of cosmological parameters on halo substructures]{Zentner03v598,Dooley14v786}; our obtained values for the SHMF slope from Illustris-Dark ($a = -0.91~ \rm{ to} -0.97$) are comparable to those found by \cite{Gao11v410} and \cite{BK09v398} ($a=0.97$ and $a=-0.935$, respectively) but steeper than e.g. \cite{Wu13v767}, who found $a=-0.865$ for a sample of massive cluster-size haloes at slightly worse resolution than Illustris-Dark's. 
This discussion lends credibility to the robustness of the results from our DMO simulation of reference and proves that the quantification of the relative comparison between Illustris-Dark and Illustris is well posed\footnote{{\sc subfind} has been noted to under-estimate the mass of subhaloes at the high-mass end near the drop-off, leading to an overestimation of the steepness \citep{Bosch16v458}. We do not consider this as a problem for our analysis, as we are mainly focused on the relative comparison between Illustris and Illustris-Dark and not on accurately extract fitting function parameters.}.

We find that, while the power-law slope of Illustris-Dark at $z=0$ across host masses is consistent with previous $N$-body findings, Illustris is characterized by a systematically shallower power-law slope ($a\approx 0.74$ at $z=0$). This change in power-law slope can also be observed at infall from Table \ref{table:fit}.
We note that a direct comparison of the other fitting parameters e.g. $\tilde \mu_1$ between Illustris and Illustris-Dark is hard, because of the result of the large change in power-law slope between the two.

\subsection{Effects of Baryons on the average abundances}
\label{sec:baryons}

To highlight the effects of baryons on the abundance of subhaloes, in the lower panels of Figure \ref{fig:shmf}, we plot the mean ratio of the SHMF in Illustris over that of Illustris-Dark. The solid curves are simply the ratio of the curves from the upper panels. The dotted curves, on the other hand, denote a similar ratio, but for the subhalo abundances considering only {\it matched host haloes} in Illustris and Illustris-Dark (see Section \ref{subsec:matching} for details on the matching procedure): here, the fractional subhalo mass $\mu$ is normalized by the respective host mass in Illustris and Illustris-Dark, but the haloes in each mass bin is chosen by the matched Illustris host mass. In practice, the dotted curves return the difference in subhalo abundance for the {\it same} host in Illustris and Illustris-Dark, removing the contribution to the solid curves from the possible changes in host masses between Illustris and Illustris-Dark. 

We find that baryonic physics affects the subhalo abundances both at infall and at the current epoch, although the magnitude of the overall effect is larger at $z=0$ than at infall.
Both at infall and $z=0$, baryons change the slope of the SHMF in Illustris, resulting in an overall reduction of the subhalo abundance at small $\mu$; however, the effect is not monotonic and the Illustris SHMF is higher for relatively larger subhaloes. 
For subhaloes of mass $m_{\rm sub} \lesssim 10^{10} \msun$, the suppression due to baryons can be as large as 50 per cent; above this transition mass, the abundance of more massive subhaloes on the other hand can be increased by as much as 60 per cent in Illustris compared to Illustris-Dark. The overall relative magnitude of the effect of baryons on the current abundances is larger for progressively more massive host haloes. 

In Appendix \ref{sec:appendix} (Figure \ref{fig:shmfres}), we show that the qualitative nature of these statements holds across resolutions, and that the quantitative baryonic effects can be considered converged across essentially all the mass ranges studied here.

\begin{figure*}
  \centering
  \includegraphics[width=\textwidth]{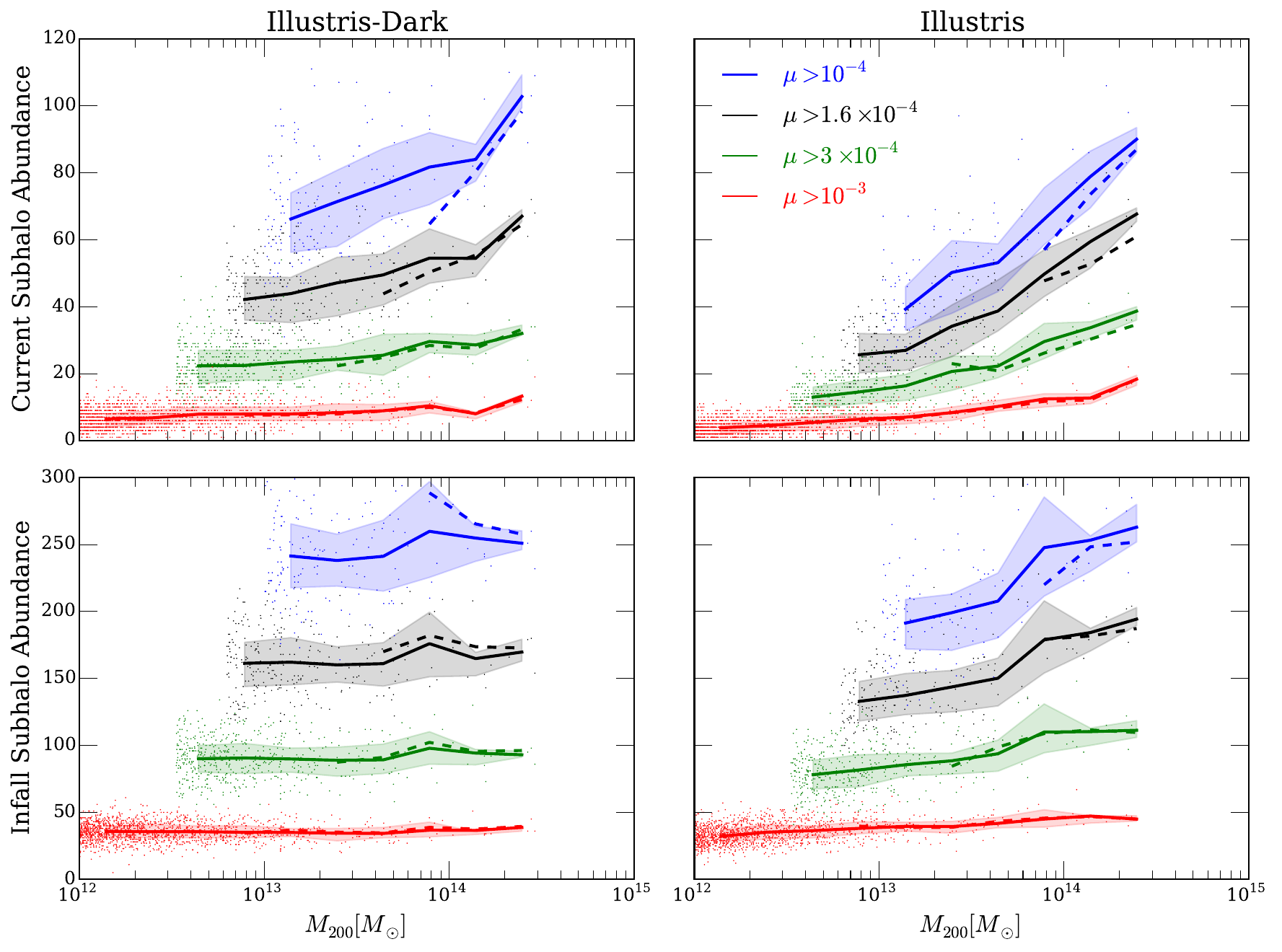}
  \caption{
  	Subhalo abundances in two cosmological simulations. The mean current (top row) and infall (bottom row) cumulative subhalo abundances are given as a function of the host halo mass for different minimum values of $\mu$.
	The colour scheme is the same for all panels, with the shaded region denoting the 25th to 75th percentile of the distributions.
	Lower resolution results from Illustris-2 and Illustris-Dark-2 are shown as dashed lines. 
	We find that only the mean infall subhalo abundance in Illustris-Dark is independent of the host halo mass, and that baryons produce a strong dependence of subhalo abundances on halo mass already at infall.
}
\label{fig:shmf_m200}
\end{figure*}
\subsection{On Self-Similarity}

We recast the subhalo abundances in Illustris and Illustris-Dark in Figure \ref{fig:shmf_m200}, to emphasize their dependencies on host halo masses. The cumulative number of subhaloes above a given normalized subhalo mass ($\mu$ and $\mu_{\rm acc}$) is given for each host in the simulation as data points, with the solid curve denoting the mean number at fixed host mass (in bins in host mass) and shaded areas encompassing the 25th and 75th percentiles. The four panels refer to results from Illustris (right) and Illustris-Dark (left), at $z=0$ (top) and infall (bottom), respectively.

First of all, inspection of the four panels of Figure \ref{fig:shmf_m200} facilitate quantification of results already mentioned above: infall abundances are larger at infall than at the current epoch (bottom vs. top rows), and baryonic effects suppress the overall subhalo abundances (left vs. right); the latter occurring already at infall. Finally, Figure \ref{fig:shmf_m200} allows us to comment on the universality of the SHMF.

The universality of the subhalo abundances has been invoked by early numerical DMO experiments \citep[e.g.][]{Moore99v524,Helmi01v323,DeLucia04v348,Diemand04v352} as a necessary reflection of the scale-free nature of gravity. The gravitational scale-free collapse would lead to the formation of self-similar haloes across masses, and indirectly to normalized subhalo abundances which are  independent of host mass and of any other host physical property (and hence exhibiting horizontal curves in plots like in Figure \ref{fig:shmf_m200}, with no scatter).
Instead, here we find that only the infall Illustris-Dark subhalo abundance is self-similar; the current Illustris-Dark SHMF is not entirely self-similar across haloes masses, since the mean subhalo abundance exhibits a weak trend with host mass; and finally, baryonic processes heavily enhance the breaking of self-similarity both at $z=0$ and infall (right panels in Figure \ref{fig:shmf_m200}).

The breaking of self-similarity in the evolved DMO SHMF had been noted already by \cite{Gao04v355}, on the grounds of the mass dependency of halo merging histories and formation times. Now, we can better understand our infall results using analytic work by JB16 who modeled the accretion of subhaloes using the merger tree algorithm developed by \cite{Parkinson08v383}.
JB16 found that the infall SHMF is universal and independent of host mass and formation time. However, self-similarity across host masses is broken when considering the current SHMF due to the evolution of subhaloes after accretion by mechanisms other than gravitational collapse.
In hierarchical structure formation, more massive haloes form and accrete subhaloes later and are hence dynamically younger compared to their less massive counterparts \citep{Navarro97v490}. As such, there is less time for the subhaloes to be tidally stripped. Additionally, massive haloes are also less concentrated and thus have weaker tidal fields. 

Our DMO results are consistent with other $N$-body simulations \citep[e.g.][]{Gao11v410} as well as analytical results \citep[e.g.][]{Bosch05v359,Jiang16v458}; however, other work such as those of \cite{DeLucia04v348} and \cite{Dolag09v399} reported no such trends with halo mass in their $N$-body simulations of cluster-size haloes (with the discrepancy likely due to the small range of host masses that were simulated and the large halo-to-halo scatter, both of which may mask the weak mass trend).  

\section{Halo-to-Halo variation of the Subhalo Abundances}
\label{sec:scatter}

\begin{figure*}
  \centering 
	\includegraphics[width=0.47\textwidth]{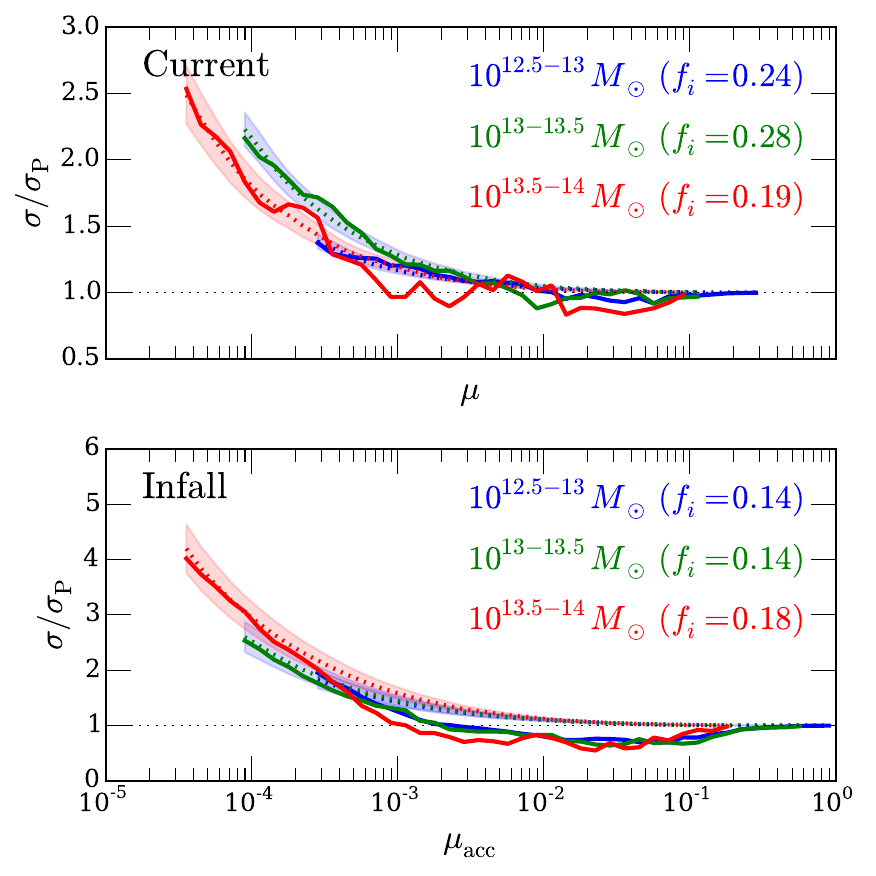}
	\includegraphics[width=0.47\textwidth]{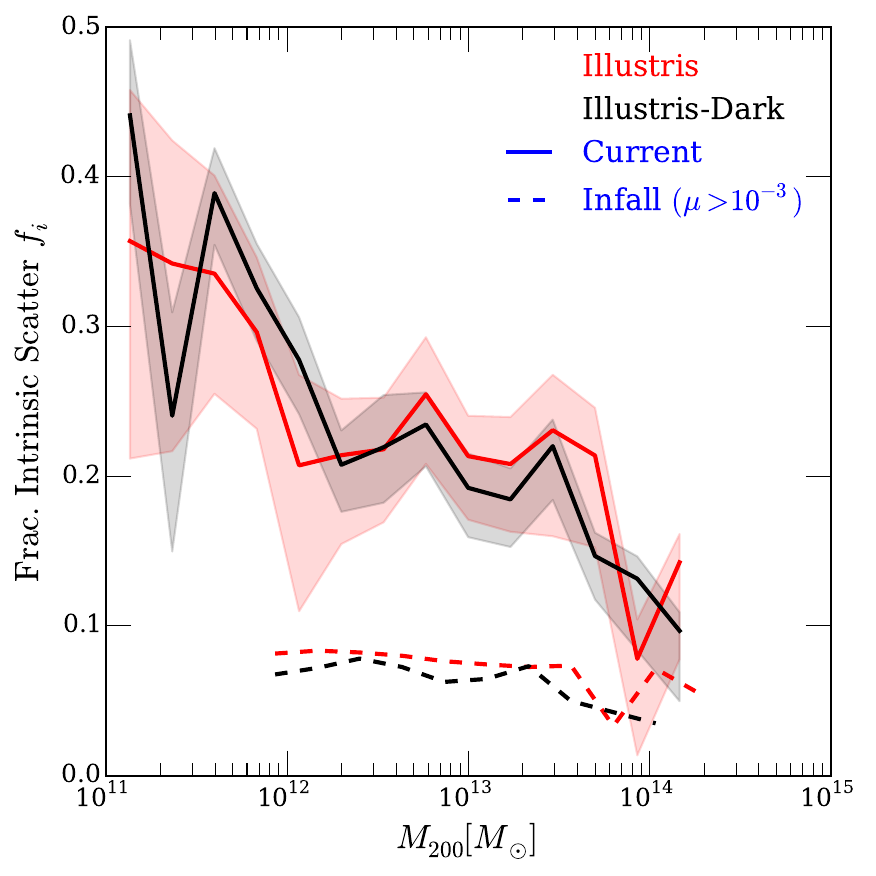}
  \caption{
  Halo-to-halo variation and intrinsic scatter in the subhalo abundances. Left: Ratio of the 1-$\sigma$ standard deviation in the SHMF over the expected Poisson error $\sigma_p$ as a function of $\mu$ (current) and $\mu_{\rm acc}$ (infall) in Illustris.
Here, coloured solid lines represent host halo masses of $10^{[12.5,13]}\,M_\odot$, $10^{[13,13.5]}\,M_\odot$ and $10^{[14,14.5]}\, M_\odot$, in blue, green and red respectively.
The dashed lines show the expected curves from models including normalized intrinsic scatter $f_i = \sigma_i/\left<N\right>$ as indicated in the legends. Intrinsic scatter increases for progressively less massive subhaloes. This plot demonstrates that Equations \ref{eq:scatter} and \ref{eq:iscatter} provide a good model for the scatter at current time in Illustris.
Right: The mean normalized intrinsic scatter $f_i$ in the SHMF as a function of the host halo mass $M_{200}$ in both Illustris (red) and Illustris-Dark (back).
Solid lines corresponds to results at current time, obtained from a fit of $\sigma$ vs. $\mu$ using equation \ref{eq:scatter}.
Shaded region at current time indicates the 1-$\sigma$ scatter obtained using a bootstrap analysis.
Dashed lines show $f_i$ at infall, for a specific cut $\mu_{\rm acc}>10^{-3}$.
The normalized intrinsic scatter at current time ($z=0$) is generally larger than at infall but also decreases with increasing $M_{200}$.
}
\label{fig:scatter}
\end{figure*}

We have thus far focused on the {\it average} abundance of subhaloes in different mass hosts. We turn now the attention to the halo-to-halo variation in the SHMFs. In Figures \ref{fig:shmf} and \ref{fig:shmf_m200}, this has been indicated as errorbars or shaded areas, denoting either the 1-$\sigma$ scatter around the mean or the first and third quartiles of the subhalo number distributions in bins of host mass. For example, for $\sim 10^{13} \msun$ hosts, the current number of subhaloes more massive than $\sim 10^{9} \msun$ can vary between about 50 and 70 in Illustris-Dark and about 35 and 45 in Illustris.

At the low-mass end of the SHMF ($\mu \lesssim 10^{-4}$), \cite{BK10v406} have found with the MS-II that the number distribution of subhaloes or the subhalo occupation distribution is no longer Poissonian but can instead be well approximated as a broader negative binomial distribution. With this model, the overall scatter in the SHMF is a result of an additional contribution from an intrinsic scatter $\sigma_{\rm i} \propto \left<N\right>$ to the variance:
\begin{equation}
\sigma^2 = \sigma^2_{\rm p} + \sigma^2_{\rm i}
\label{eq:scatter}
\end{equation}
where $\sigma^2$ is the mean-square-error in the SHMF and $\sigma^2_{\rm p} = \left<N\right> $ is the contribution from Poisson fluctuations.

The intrinsic scatter $\sigma_i$ is related to the reduced second moment of 
the subhalo occupation distribution
$\mathcal M_2 \equiv \left< N \left(N-1 \right)\right> ^{1/2} / \left< N \right>$  through
\begin{equation}
f_i \equiv \frac{\sigma_i}{\left< N \right>} = \sqrt{\mathcal M_2^2 -1}
\label{eq:iscatter}
\end{equation}
where we have defined $f_i$ to be the intrinsic scatter of the distribution normalized by the subhalo abundance \citep{Bosch05v359,BK10v406}.
$\mathcal M_2$ and other higher moments $\mathcal M_{3,4,...}$ describe the deviation of the distribution from Poisson and are typically used in studies involving halo occupation distribution (HOD) models \citep[e.g.][]{Seljak00v318,Ma00v543,Peacock00v318}.
Distributions that are narrower and broader than Poisson have reduced moments $\mathcal M<1$ and $\mathcal M>1$ respectively.
In the negative binomial model of subhalo occupation statistics, $\mathcal M_2 > 1$ and is hence super-Poissonian.
With the assumption that $\sigma_i \propto \left< N\right>$, then $f_i$ is expected to be independent of $\left< N \right>$ and $\mu$ for a fixed halo mass.

We follow this formalism and apply Equations \ref{eq:scatter} and \ref{eq:iscatter} to both Illustris-Dark and Illustris in order to quantify the magnitude of the normalized intrinsic scatter $f_i$, under the assumption that, if Equation \ref{eq:cdf} well describes the mean shape of the SHMF in both Illustris and Illustris-Dark, Equation \ref{eq:scatter} may well describe the halo-to-halo variation in both too. 

To show that the intrinsic scatter model works in Illustris, we plot the ratio of the 1-$\sigma$ halo-to-halo scatter over the expected Poisson error $\sigma_p$ in Figure \ref{fig:scatter} as a function of $\mu$ (left panels). We show this for haloes in three different mass ranges using both the current (top) and infall (bottom) abundances, with solid curves.  
We find that the scatter becomes significantly super-Poissonian for $\mu \lesssim 10^{-3}$, leading to a broader subhalo distribution. 
$\sigma/\sigma_p$ (Figure \ref{fig:scatter}, left panel) is larger at infall than at $z=0$, hence the intrinsic scatter contributes more to the total scatter at infall compared to $z=0$. In other words, the subhhalo distribution is more Poissonian at the current epoch compared to infall.

In the same panels (left panels of Figure \ref{fig:scatter}), the dotted curves show models that include normalized intrinsic scatter with values indicated in the legend, while the shaded regions correspond to varying $f_i$ by $\pm 0.2$. The agreement between the solid and dotted curves demonstrates that the intrinsic scatter model where $f_i$ is independent of $\mu$ adequately describes subhalo occupation distribution (for $\mu \lesssim 10^{-2}$) also in a baryonic physics run like Illustris for the current epoch $z=0$.
We hence extend this procedure to other halo masses at $z=0$, using a least-squares analysis to determine $f_i$ by fitting equation \ref{eq:scatter} to a plot of $\sigma$ vs. $\mu$ or equivalently, $\sigma$ vs. $\left< N \right>$. In addition, we also perform a bootstrap analysis to estimate the scatter in $f_i$ at each halo mass.

From the left panels of Figure \ref{fig:scatter}, we also find both the current and infall scatters to be sub-Poissonian for  $\mu,\mu_{\rm acc}\gtrsim 10^{-2}$: this is not captured by the intrinsic scatter model of \cite{BK10v406}. At fixed halo mass, the sub-Poissonity reduces between infall to the current epoch.
Nonetheless, we have plotted the best-fit intrinsic scatter model to illustrate the best-fit normalized intrinsic scatter $f_i$ that can be attained for $z=0$.
For infall, the significant sub-Poissonity results in fits that are not entirely reliable. Hence, we have chosen to avoid extending the fitting procedure to all halo masses at infall and instead measure $f_i$ using specific $\mu_{\rm acc}$ cuts. 
For brevity, we show infall results only for $\mu_{\rm acc}>10^{-3}$, although the following conclusions also hold for other $\mu_{\rm acc}$ cuts.

The right panel of Figure \ref{fig:scatter} shows the mean best-fit normalized intrinsic scatter $f_i$ plotted as a function of host mass, for Illustris (red) and Illustris-Dark (black) at both $z=0$ (solid curves) and infall (dashed curves).
We note again that results for $z=0$ are obtained from a fit of $\sigma$ vs. $\mu$ to equation \ref{eq:scatter}, with shaded regions indicating the standard deviation arising from our bootstrap analysis. For the infall curves, our choice of $\mu_{\rm acc}>10^{-3}$ restricts the host mass range to $M_{200} > 10^{12}\,M_\odot$ by the resolution limit.

Four main results emerge:
1) the inclusion of baryons in Illustris does not appreciably affect the halo-to-halo variance with respect to Illustris-Dark, this being the case both at infall and at $z=0$; 
2) the subhalo abundance is super-Poissonian for $\mu \lesssim 10^{-3}$ and sub-Poissonian for  $\mu \gtrsim 10^{-3}$, with deviations from Poissonity decreasing from infall to $z=0$;
3) the normalized intrinsic scatter increases between infall and $z=0$; 
and 4) in the evolved subhalo abundances, the normalized intrinsic scatter decreases with increasing host mass, decreasing from 40 per cent for $10^{11}\,M_\odot$ hosts to 10 per cent for $10^{14}\,M_\odot$ hosts. The last trend is, in comparison, essentially inexistent at infall.

Results \#1 and \#2 suggest that the main source of non-Poissonian scatter in subhalo occupation comes from host properties that are independent of baryonic physics, as can be the case for the diverse accretion histories of the hosts, whose trends are controlled by the overall hierarchical growth dominated by the DM component. 
This would explain why non-Poissonity is present already at infall even in Illustris-Dark.  Our result for the sub-Poissonity in halo occupation statistics is in agreement with the findings of \cite{Jiang2016}, who argue that the weaker non-Poissonity at $z=0$ is a result of host-subhalo interactions after accretion. Their argument explains results from the Bolshoi and MultiDark simulations, according to which the subhalo abundance of early-forming haloes is more Poissonian than those of late-forming ones and that the sub-Poissonity can be described by a small modification to equation \ref{eq:scatter}.
The super-Poissonity for $\mu \lesssim 10^{-3}$ has also been observed in other previous N-body studies \cite[e.g.][]{BK10v406,Busha11v743b,Wu13v767}. On the other hand, the semi-analytic model of \cite{Bosch05v359} fails to reproduce the non-Poissonian characteristics in subhalo occupation distribution as a result of the extended Press-Schecter formalism used to derive the merger trees.

Interestingly, our result \#3 shows that when normalized by the subhalo abundance, the normalized intrinsic scatter $f_i$ increases between infall and $z=0$, despite its smaller contribution to the total halo-to-halo scatter (result \#2). 
This is a result of the uneven change in $f_i$ and subhalo abundance $\left< N \right>$ from infall to $z=0$, thus affecting the ratio of the intrinsic and Poisson scatters $\sigma_{\rm i}/\sigma_{\rm p} = f_i\sqrt{\left< N \right>}$.

We expand on the physical origin of the scatter in Section \ref{sec:props_scatter} and spell out here a remarkable quantitative conclusion: for Milky Way-like haloes, the normalized intrinsic scatter increases from about eight per cent at infall to about 25 per cent at the current epoch, across subhalo masses; these correspond to abundances of subhaloes more massive than $10^9\msun$ in the range $4.05\pm2.23$ and $6.59\pm2.93$ for full-physics and DMO Milky Way haloes respectively, within 1-$\sigma$.

Finally, our result \#4 on the host mass dependence of the normalized intrinsic scatter has been quite debated over the literature, for the DMO case. We speculate it is the evidence that more massive objects tend to form in more uniform environments than lower mass hosts: hierarchical structure formation indeed acts so that massive objects are biased to form in regions of higher density contrast while low-mass objects can form in both high and low density contrast regions \citep{Busha11v743}. This, in addition to the fact that the larger numbers of low-mass haloes enable them to sample their environments more effectively, may be the reason for larger normalized intrinsic scatter towards the low-mass end of the host population. With the Bolshoi simulation, \cite{Busha11v743} also found a trend in halo mass with the scatter of the $z=0$ subhalo velocity function, in agreement with our results from Figure \ref{fig:scatter} but in contrast with \cite{BK10v406}, who suggest from the MS-II that the intrinsic scatter is independent of halo mass. 
\cite{Busha11v743} speculated the difference between these results to be due to the larger box size of the Bolshoi simulation ($250h^{-1}$ Mpc) compared to MS-II (137 Mpc), and hence attributing to Bolshoi an enhanced capability of sampling a wider range of environments. While we cannot agree with this solution (given that the Illustris-Dark box size is closer to MS-II's rather than Bolshoi's), we have independently repeated the analysis on the Bolshoi simulation using the {\sc rockstar} halo finder \citep{Behroozi13v762} and in terms of cumulative subhalo velocity function, and confirm \cite{Busha11v743}'s conclusions.

\begin{figure*} 
  \centering 
        \includegraphics[width=0.45\textwidth]{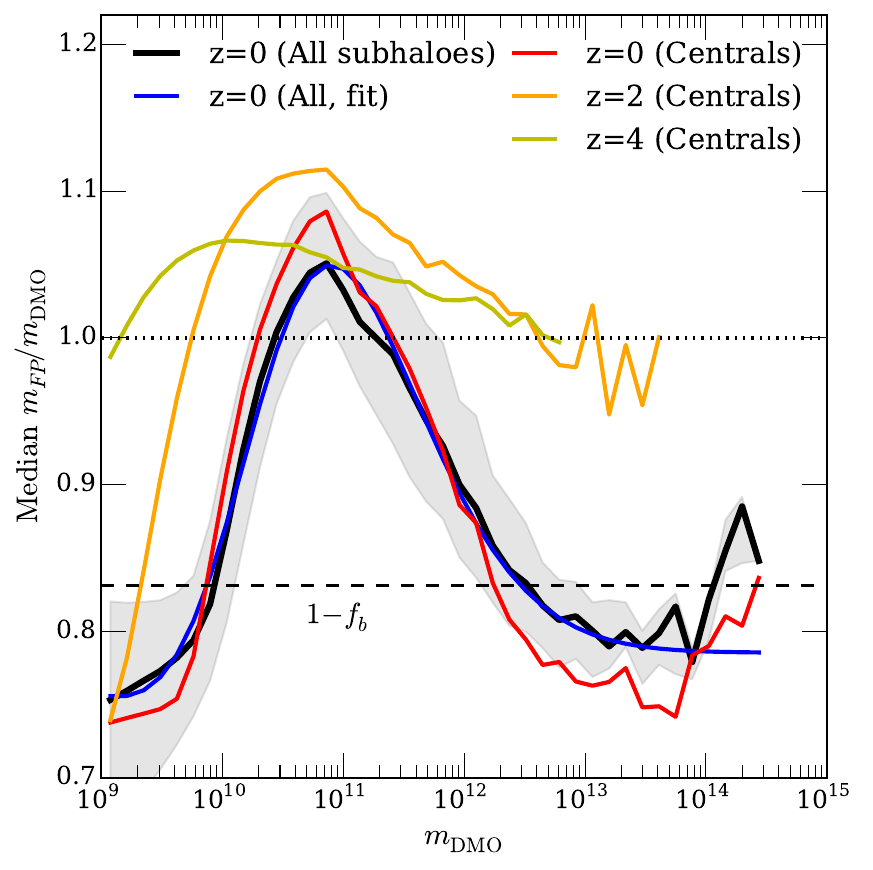}
	\includegraphics[width=0.455\textwidth]{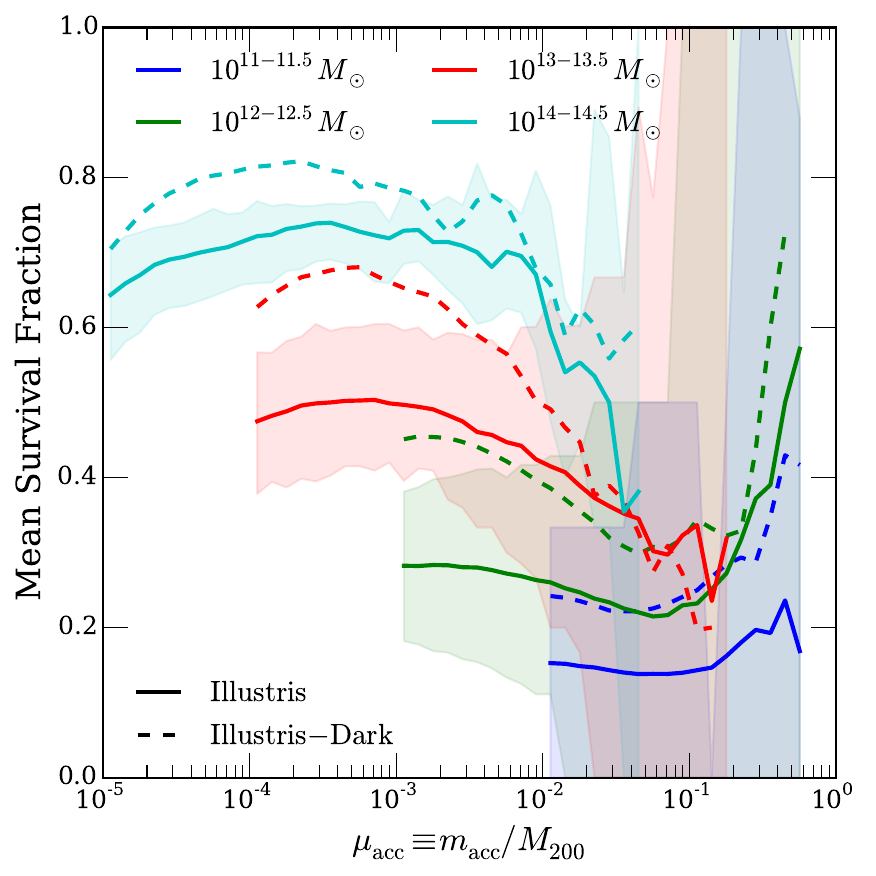}
	\caption{Left: Black line: Median ratio of subhalo mass in Illustris to matched subhaloes in Illustris-Dark as a function of the Illustris-Dark mass at $z=0$.
	Shaded area denotes the 25th to 75th percentile in the distributions while the blue line shows the best fit curve to equation \ref{eq:corr}.
	The red, orange and yellow lines show the median ratio of halo masses (FOF groups) in Illustris to Illustris-Dark at $z=0$, 2 and 4 respectively. 
	Horizontal dashed lines denote the fraction $1-f_b \equiv (\Omega_m-\Omega_b)/\Omega_m$.
	Right: The mean survival fraction of accreted subhaloes as a function of the normalized subhalo mass at accretion. 
	Results for shown for matched Illustris and Illustris-Dark haloes in solid and dashed lines respectively,
	with the shaded region showing the 25th to 75th percentile of the distributions in Illustris.
	The survival fraction varies significantly with host mass.
	Note also the effect of baryons in suppressing the current time to infall abundance ratio in lighter subhaloes (small $\mu$).
	}
\label{fig:massratio_and_survivalfraction}
\end{figure*}
\section{Understanding the Effects of Baryons on the average subhalo abundances}
\label{sec:baryons2}

In the previous sections (Figures \ref{fig:shmf} and \ref{fig:shmf_m200}), we have demonstrated that the inclusion of baryonic physics tilts the $z=0$ cumulative SHMFs for host masses in the range $10^{11} - 10^{14.5}\msun$: namely, baryons reduce the abundance of small subhaloes ($\lesssim 10^{10} \msun$) compared to the DMO case and make the SHMF shallower. We have also seen that baryonic effects are manifest already at infall through the unevolved SHMF, and that low-mass isolated haloes host relatively less subhaloes than more massive hosts compared to the DMO case. In this section, we are going to further quantify these effects and identify their physical reasons.

There are two possibilities for which, at any given time, the SHMFs from Illustris and Illustris-Dark can differ: 1) because the masses of subhalo/hosts are different in the two runs at any give time and so they shift the values of $\mu$ on the x-axis of e.g. Figure \ref{fig:shmf}; 2) because the actual number of subhaloes in the two cases are different, therefore shifting up and down the SHMF curves at different subhalo masses. 
In fact, the reduction of a subhalo mass can be so severe in Illustris vs. Illustris-Dark, or vice versa, that a subhalo might not pass the resolution mass threshold in one of the two runs, hence contributing to the latter effect.
By keeping this in mind, we want to distinguish between the two possibilities. 

In the left panel of Figure \ref{fig:massratio_and_survivalfraction}, we quantify how baryonic physics alter the {\it mass} of subhaloes at $z=0$ (black and blue solid curves) and host haloes at various redshifts (coloured solid curves). The solid curves represent the median ratio between the total masses of {\it matched} (sub)haloes, as a function of the DMO mass. At $z=0$, the subhalo masses in Illustris are generally suppressed relatively to Illustris-Dark, except for objects of about $10^{11} M_\odot$, where the ratio peaks and the subhaloes in Illustris have larger masses than their DMO analogs. Similarly it happens for isolated haloes, at both the current epoch as well as at higher redshifts: baryons imprint a non-monotonic effect on the masses of haloes, even in isolation i.e. even before they may become satellites of larger objects. The masses of isolated haloes at low redshifts are smaller in the FP run compared to the DMO run, in agreement with what was already shown in \cite{Vogelsberger14v444}. This finding justifies why the SHMFs in the two runs are different already at infall: the infall SHMF is a complex quantity that depends on both the mass function of isolated haloes (or `centrals') and the infall time when they are accreted; the masses of isolated haloes are different with baryonic physics for redshifts $z=0$, 2 and 4, shown as possible subhalo accretion times. \\

The decrease in the total subhalo abundance in Illustris can also be characterised by quantifying the survival fraction, as shown in the right panel of Figure \ref{fig:massratio_and_survivalfraction}. The survival fraction is calculated as the ratio of all accreted subhaloes (above a given $\mu$) that can still be identified in the final snapshot at $z=0$, above the mass threshold of $10^9 M_\odot$ (see Section \ref{subsec:limits}). In Figure \ref{fig:massratio_and_survivalfraction}, we show the survival fractions of accreted subhaloes in Illustris (solid) and Illustris-Dark (dashed), as a function of the normalized accretion mass and after having matched the host haloes between the two runs. 
The shaded areas show the 25th to 75th percentile of the distributions in Illustris.
First of all, clearly, many subhaloes that are accreted do not survive to the present but are instead disrupted.
In each run, the survival fraction increases as a function of halo mass: this is due to the fact that massive haloes are dynamically younger and hence accreted subhaloes have less time on average to be disrupted.  In cluster-sized haloes ($10^{14} M_\odot$), about 60 per cent of subhaloes with $\mu_{\rm acc} > 10^{-2}$ survive to the present. The fraction drops to about 40 per cent and 20 per cent in $10^{13} M_\odot$ and $10^{12} M_\odot$ haloes, highlighting the additional stripping that takes place in these small hosts. Moreover, the survival rates of subhaloes accreted into a host of a given mass depends on their mass at accretion. 

More importantly, for the purposes of this paper, the survival fraction generally decreases going from Illustris-Dark to Illustris, indicating that accreted subhaloes of a given mass ratio in baryonic runs are less likely to survive.\\

\begin{figure}
  \centering
	\includegraphics[width=0.47\textwidth]{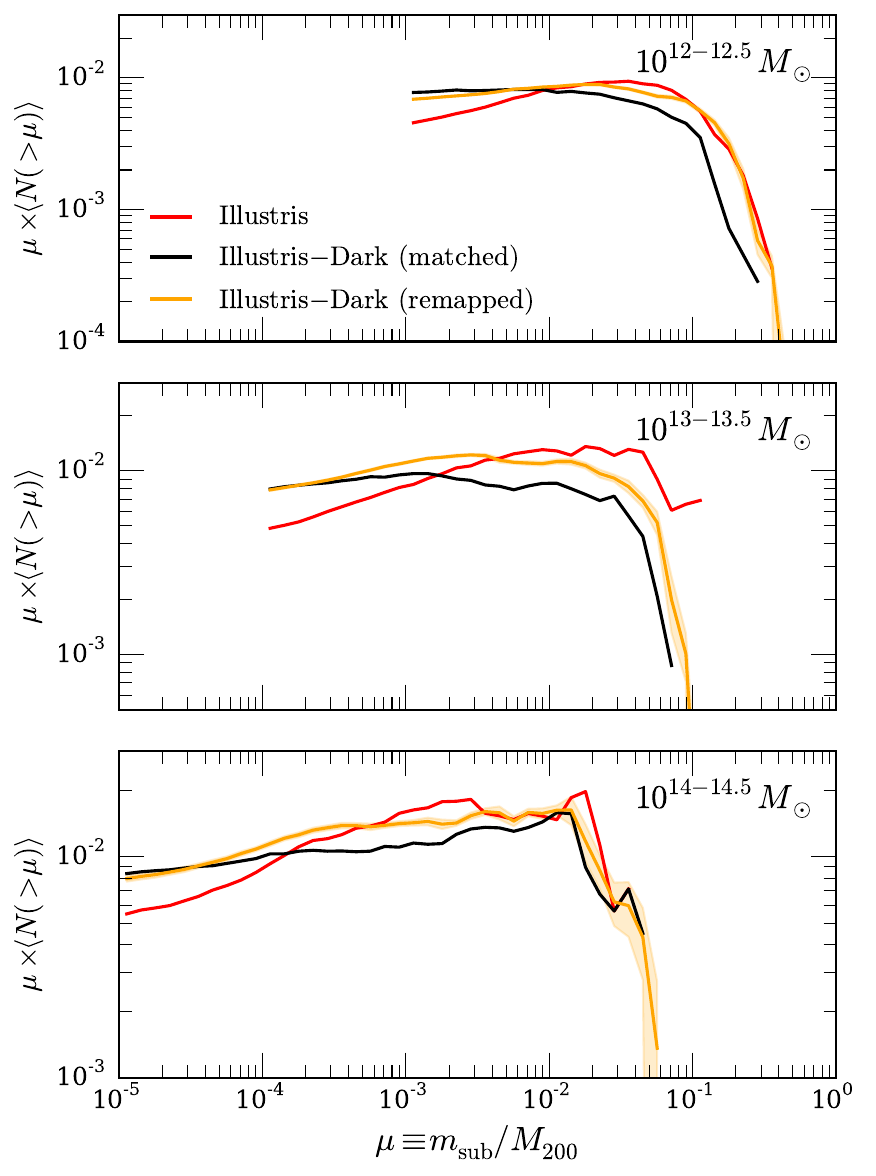}
	\caption{
	The mean current subhalo mass functions (scaled by $\mu$), showing the result of mapping Illustris-Dark subhalo masses into Illustris subhalo masses following the mass correction from equation \ref{eq:corr}. Results are shown here for matched haloes only, with haloes chosen according to the mass or matched mass in Illustris.
	Orange curves show the remapped SHMF from Illustris-Dark while red and black curves correspond to the SHMF from Illustris and Illustris-Dark.
	The remapping procedure fails to correct the low-mass end but improves the agreement at the high-mass end with Illustris.
  }
\label{fig:shmf_corrected}
\end{figure}

The final $z=0$ SHMF in Illustris compared to Illustris-Dark is therefore the complex result of a) a modified mass spectrum of accreted subhaloes; and b) a different efficiency of Illustris subhaloes to survive within the potential wells of Illustris host haloes. The modified mass spectrum of $z=0$ subhaloes of Figure \ref{fig:massratio_and_survivalfraction} (left panel) can be fit with an equation of the form:
\begin{equation}
r(\lg m) =r_0+ A \exp{\left[\frac{-x^2}{2b^2}\right]} \left[w + \Phi \left(\alpha x \right) \right]
,\, x\equiv \lg m - m_0
\label{eq:corr}
\end{equation}
where $\Phi$ is the error function and $A,m_0, b,w,\alpha$ and $r_0$ are fitting parameters (blue thin solid curve)\footnote{
We find the following values to provide an excellent fit, reproducing the peak, width, asymmetry and normalization of the curve well:
$A=0.23$, $m_0=10.3$, $b=1.06$, $\alpha=1.15$, $f=0.70$, $r_0=0.79$.}.
In Figure \ref{fig:shmf_corrected}, we show how the Illustris-Dark SHMF would be modified if we map the Illustris-Dark subhalo masses into Illustris subhalo masses following the mass correction from equation \ref{eq:corr}, for matched hosts of different mass. In addition, the host masses are corrected by using the host mass of the matched Illustris analog. The resultant {\it remapped} SHMFs are shown as orange curves, taking into account the scatter of Figure \ref{fig:massratio_and_survivalfraction} left panel, which we assumed to be $\sigma=0.1$ independent of host mass. These are compared to the Illustris and Illustris-Dark SHMFs in red and black solid curves, for matched hosts only, respectively. 
It is evident that the remapped SHMF fails to modify the low-mass end, continuing to over-estimate the Illustris abundance. 
At the high-mass end, agreement between Illustris and the remapped curves is much improved compared to Illustris-Dark. 
This confirms that the physical mechanisms which govern the host-subhalo interaction indeed affect the survivability of Illustris subhaloes  in a non negligible and diversified manner in comparison to the Illustris-Dark case. \\

We note that the Illustris results presented here are specific to our underlying galaxy formation model and that simulations with differing feedback mechanisms may produce different results, both quantitatively and qualitatively. Our results in Figure \ref{fig:shmf} are in perfect agreement with the analysis presented in \cite{Despali2016}, who had analyzed both the Illustris and EAGLE cosmological simulations, although only at $z=0$. There, they found that the Illustris and EAGLE simulations differ at the high-mass end of the SHMF, even though both predict a suppression of small substructures. In EAGLE, the relative discrepancy between the hydrodynamic and DMO runs does not increase above unity for subhaloes more massive than $10^{10} \msun$. The difference is attributed to the different AGN feedback implementation: the weaker feedback in EAGLE is unable to expel enough baryons to suppress the halo mass. In a similar vein, results depicted in Figure \ref{fig:massratio_and_survivalfraction} are different between Illustris and EAGLE: in EAGLE, the relative hydro-to-DMO mass increases almost monotonically (and can be fitted with a double sigmoidal function) from 0.7 to 1.0 for subhaloes in the same mass range \citep{Schaller15v451}. A similar trend is obtained by \cite{Sawala13v431} for the GIMIC cosmological simulation. However, the lack of AGN feedback in GIMIC resulted in a single sigmoidal function describing the relative hydro-to-DMO masses \citep{Schaller15v451}. 

Comparing our results to recent zoom-in simulations, we find instead similarity with the outcome of a MW-like halo analyzed in \cite{Zhu16v458}, a hydrodynamic analog of one of the haloes of the Aquarius Project, also simulated with {\sc arepo} \citep{Aquarius} but with a slightly different galaxy physics model \citep{Marinacci14v437}. The authors found a 50 per cent reduction of the final subhalo abundance in the hydro run compared to the DMO run down to a subhalo mass of $3 \times 10^6 \msun$. At the same time, the abundance of subhaloes with $v_{\rm max} > 35\,{\rm km\,s^{-1}}$ is increased in the hydro run relative to the DMO. These results confirm those from our analysis (for MW-like haloes) and extends them to lower subhalo masses. 

\section{Physical Origin of Baryonic Effects and Scatter}

In this section, we tackle the following questions: 1) What are the physical mechanisms which modifies the mass of subhaloes and hosts differently in Illustris vs. Illustris-Dark? 2) How do different subhalo-host interactions affect the survivability of subhaloes in the two runs?

In relation to the unevolved SHMFs, Illustris halo masses are modified in different manners at different redshifts with respect to the DMO case: this makes a systematic study of the underlying physical processes very challenging. We postpone this task to a later dedicated work, while here we simply speculate on possible explanations. First of all, in no case the (sub)halo mass variations of Figure \ref{fig:massratio_and_survivalfraction} left panel can be justified by a change by $\frac{\Omega_m - \Omega_b}{\Omega_m}$ of the DM particle mass between the baryonic and DMO runs (see horizontal dashed line in Figure \ref{fig:massratio_and_survivalfraction}). Yet, it may happen that galactic winds and photoionization from a UV background may induce gas expulsion, which together with a back-reaction and redistribution of DM, cause a mass suppression in Illustris vs. Illustris-Dark, at least at the low-mass end and in certain redshift regimes. At intermediate masses, the presence of a central stellar and gaseous component may deepen the halo potential well, favour gas accretion and hence induce a mass enhancement in isolated objects in Illustris vs. Illustris-Dark.\\

On the other hand, the physical mechanisms acting on subhaloes {\it after their infall} into more massive hosts have been already extensively studied, specifically in the case of DMO and gravity-only simulations. As already mentioned, it has been demonstrated in gravity-only simulations that hosts that form later accrete the bulk of their subhaloes at a later time: their subhaloes orbit in the host potential for a shorter time and hence have less time to lose their mass because of (tidal) stripping, in turn having a higher chance to not get completely disrupted. On the other hand, stripping affects subhaloes by removing material from their outskirts progressively inwards: steeper matter density profiles in subhaloes enhance their resilience to stripping and total disruption; on the contrary, host haloes with steeper matter density profiles are more efficient at stripping their satellite haloes \citep[see][for a review]{Jiang16v458}. 

These phenomena have been evoked in gravity-only simulations to explain both the shape of the evolved {\it average} SHMF in general, and its scatter at fixed host mass. More massive haloes form more recently than low-mass haloes and are less concentrated: they host a larger number of subhaloes above a given normalized mass than their lower mass host counterparts \citep[e.g.][]{Navarro97v490}. On the other hand, at fixed host mass, \cite{Gao11v410} have found in MS-II that DM haloes with higher concentrations and earlier formation times contain fewer subhaloes. The mass fraction in substructure, related to the subhalo abundance, has also been found to depend on the halo shape and spin \citep{Wang11v413}; the impact of environment has been much more controversial, with some studies finding environmental dependence of the subhalo abundance \citep{Fakhouri10v401,Croft12v425,Wang11v413}, and with others finding no such evidence \citep{BK10v406,Jeeson11v415}.

In addition to (tidal and ram pressure) stripping, additional effects from galaxies can also alter subhalo abundances in hydrodynamic simulations. Specifically for disk galaxies, it has been found in single zoom-in simulations that the presence of discs can gravitionally shock subhaloes during disc passages and lead to additional subhalo destruction \citep{D'Onghia10v709,Yurin15v452}.

 \begin{figure*} 
   \centering 
 	\includegraphics[width=\textwidth]{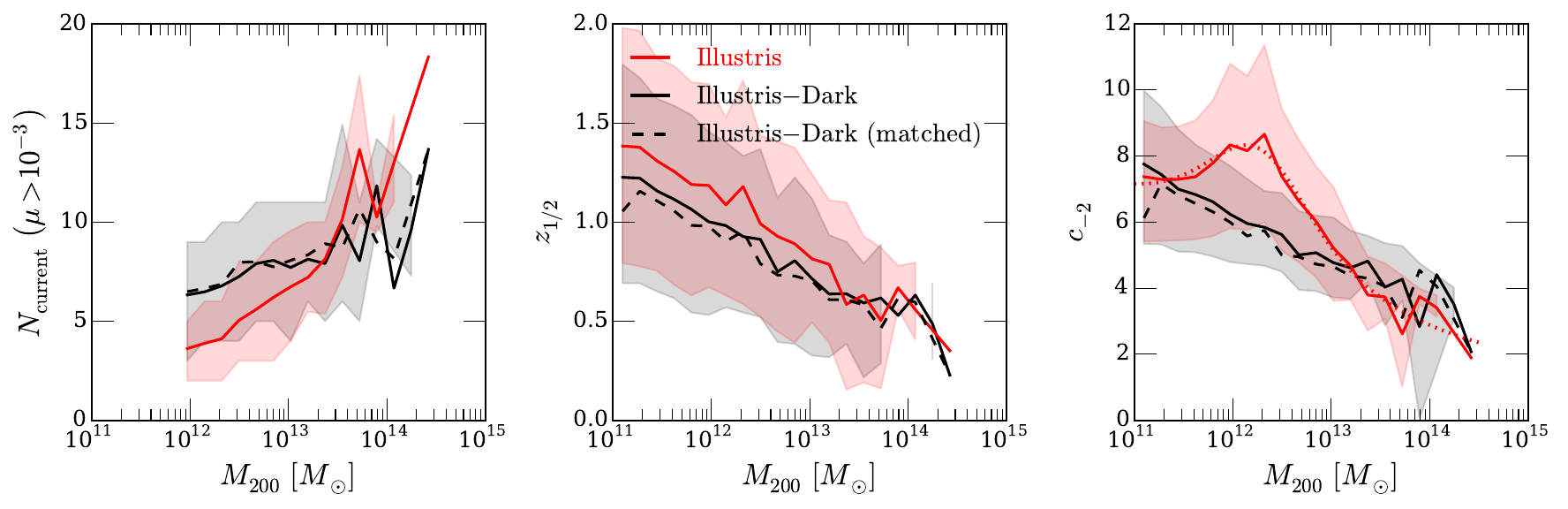}
 	\caption{Physical properties of host haloes as a function of mass in Illustris (red solid curve) and Illustris-Dark (black solid curve) at the current epoch. The black dashed curve show the Illustris-Dark results considering the matched analogs of Illustris haloes.
 		Left: abundance of subhaloes. Center: average halo formation redshift. Right: DM halo concentration. In all cases, solid curves denote mean properties at fixed halo mass, and the shaded areas indicate the 1-$\sigma$ scatter. Noticeably, the concentration-mass relation is dramatically altered by the baryonic physics in Illustris, and in a non-monotonic manner. 
 		A functional fit for the Illustris c--M relation as described by Equation \ref{eq:concmass} is shown in the right panel as red dotted curves.
 	}
 \label{fig:haloprop}
 \end{figure*}
\subsection{Effects of Host Properties on the average subhalo abundance at $z=0$}

In Figure \ref{fig:haloprop}, we show how host halo properties differ in Illustris vs. Illustris-Dark, as a function of halo mass. Here we focus on halo formation redshift (center) and concentration of the DM halo profiles (right), according to definitions of Section \ref{subsec:hostprops}. In the left panel we report the abundances of relatively massive subhaloes, exactly as in Figure \ref{fig:shmf_m200}, as a reminder that baryonic modification is non-monotonic with host mass.

While FP and DMO haloes have similar halo formation redshifts, except for a slight shift, baryonic physics in Illustris dramatically alter the concentration--mass relation of isolated haloes at $z=0$: in particular, haloes as massive as the Milky Way, or more, are more concentrated than their DMO analogs. At the low and high-mass ends probed in Illustris however, the concentrations are similar to, if not smaller than in Illustris-Dark (see Pillepich et al. in prep., for more details). 
For example, a Milky Way mass object in Illustris is more resistant to (tidal) stripping if it is a subhalo of a larger host, while, as a host halo, it induces an enhanced (tidal) stripping and mass loss in its satellites, when compared to its analog in Illustris-Dark.

In Illustris-Dark, the mean concentration--mass relation shown in Figure \ref{fig:haloprop} can be described by a power law: $c_{-2}=A \left(M_{200}/10^{14} \msun \right))^B$, where we find $A=3.534\pm0.001$ and $B=-0.1074\pm0.0004$. For Illustris, we find that the addition of a Lorentzian term can describe the resulting shape of the c--M curve. As such, we fit the following curve:
\begin{equation}
\lg c_{-2} = \lg A + B(\lg M -14) + \frac{D/(2\pi)}{(\lg M - C)^2 + (D/2)^2}
\label{eq:concmass}
\end{equation}
where the first two terms involving $A$ and $B$ reflect the DMO power law written in logarithmic form. The final term with fitting parameters $C$ and $D$ describes the non-monotonic modification by baryons. We obtain from least-squares analysis the following parameters for Illustris: $A=2.39,B=-0.121, C=12.32,D=1.93$. The resultant curve from the fit is shown in the last panel of Figure \ref{fig:haloprop} as a red dotted curve. The value of the parameter $C$ reflects the peak of the Lorentzian at $10^{12.32} \msun$, where the halo concentration and contraction are maximised.

We believe that the non-monotonic modification of the concentration--mass relation due to baryonic physics is the main reason for the non-monotonic alteration of the Illustris SHMF at $z=0$. However, this is probably not the only cause for the SHMF in Illustris to be different. In fact, such an effect may add to 1) other baryons-induced changes to the inner structures of subhaloes (which we do not consider in depth here, but which we have touched upon in Figure \ref{fig:massratio_and_survivalfraction}) and 2) the effect of ram pressure stripping. This is obviously absent in Illustris-Dark: by acting on the gaseous content of Illustris subhaloes and by depending on the physical and thermodynamical properties of the host haloes, the effectiveness of ram pressure stripping may also vary according to host and subhalo masses. These effects will be systematically studied in future work.

The baryonic contraction of haloes found here in Illustris has also been observed both in other  cosmological simulations \citep[e.g.][]{Duffy10v405} as well as  zoom-in simulations \citep[e.g.][]{Marinacci14v437,Zhu16v458,Fiacconi16v826}, although in different halo mass regimes. We stress here again that these results, especially that of the concentration--mass (c--M) relation, depend strongly on the feedback implementation of the simulation: no consensus has been reached on the magnitude of baryonic contraction in haloes. In fact, this manifests as a difference with the EAGLE simulation, which finds very similar c--M relations between the hydro and DMO runs, i.e. the c--M relation in EAGLE decreases monotonically with halo mass \citep{Schaller15v451}\footnote{In EAGLE, $A=5.70\pm0.24,B=-0.074\pm0.006$ versus $A=5.22\pm0.10,B=-0.099\pm0.003$ in the DMO run \citep{Schaller15v451}.}, in contrast to the non-monotonic behavior observed in Illustris.

\begin{figure*}
  \centering 
  \includegraphics[width=\textwidth]{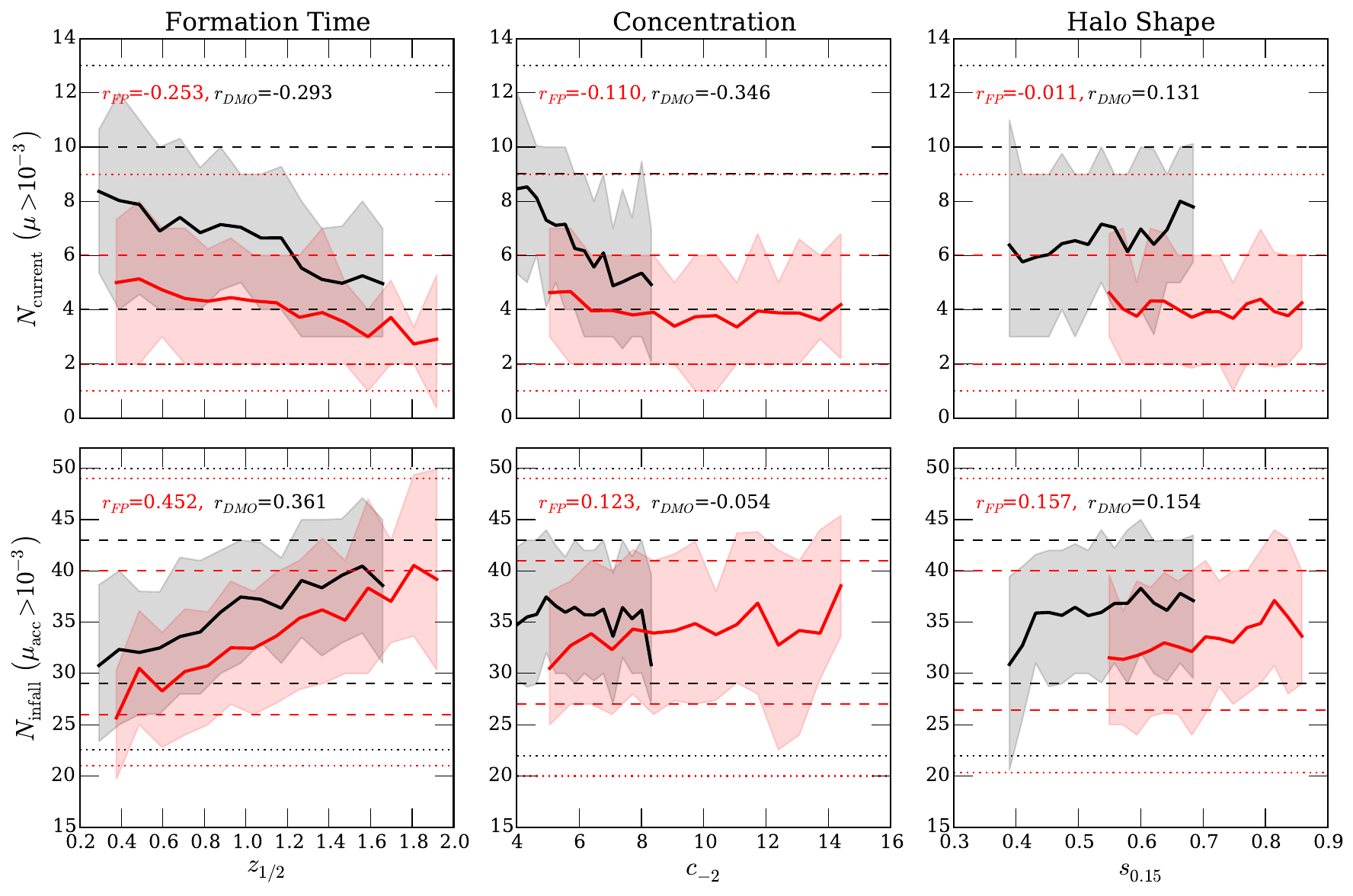}
  \caption{Physical origin of the halo-to-halo variation in subhalo abundance at fixed host mass. The current (top) and infall (bottom) abundance of subhaloes in host haloes of masses between $10^{12} -10^{12.5} M_\odot$ is plotted as a function of various host properties, in all cases at $z=0$.
	Red and black curves correspond to results from Illustris and Illustris-Dark respectively.
	We use a cutoff of $\mu > 10^{-3}$ for the current abundance and $\mu_{\rm acc} > 10^{-3}$ for the infall abundance.
	The first, second and last columns show the correlation between abundance and formation time $z_{1/2}$, concentration $c_{-2}$ and shape $s$ (last column) of the host halo, respectively.
	Dashed and dotted thin horizontal curves denote the 1 and 2-$\sigma$ scatter of subhalo abundance in the selected host mass bin. Solid curves denote mean abundance at fixed host properties within the 5th and 95th percentiles of the host property distributions.
	The text in each plot shows the calculated Pearson correlation coefficients for Illustris ($r_{\rm FP}$) and Illustris-Dark ($r_{\rm DMO}$).
}
\label{fig:propcorr}
\end{figure*}

\subsection{Origin of the halo-to-halo variation}
\label{sec:props_scatter}

We have previously noted that, even after removing the Poisson scatter, a substantial intrinsic host-to-host variation remains in both the current and infall subhalo abundances. We close this Section by examining how the SHMF scatter depends on halo formation time, concentration and DM halo shape and by extending previous $N$-body studies not only to the baryonic case at $z=0$, but also to infall.  

We begin the analysis of the scatter by studying haloes in a narrow mass range, to reduce the known correlation between halo mass and subhalo abundance. In Figure \ref{fig:propcorr}, results are shown for Milky Way-like hosts in the mass range $10^{12}$ to $10^{12.5} M_\odot$ (992 and 1052 objects in Illustris and Illustris-Dark respectively), at $z=0$ (top) and infall (bottom). At fixed halo mass, the dependence of subhalo abundance ($\mu > 10^{-3}$, i.e. larger than about $10^9M_\odot$) is given as a function of halo formation redshift, DM halo concentration and DM halo shape at $r=0.15 R_{200}$ for Illustris-Dark (black) and Illustris (red). Dashed and dotted thin horizontal lines denote the 1 and 2-$\sigma$ scatter of subhalo abundance in the selected host mass bin. Solid curves denote the mean abundance at fixed host properties within the 5th and 95th percentiles of the host property distributions. Shaded areas around the mean curves denote the variation in subhalo richness at fixed host property. The numbers reported in each panel are the Pearson correlation coefficients for each pair of subhalo abundance and host property. Although we associate subhaloes to their hosts using their 3-dimensional (3D) separations, the results in this Section hold for {\it projected} abundances as well. In general, considering 2D-projected separations would result in abundances that are systematically larger by about 50\%, regardless of halo mass and other host properties. This would introduce a normalization shift in Figure \ref{fig:propcorr}, but would not affect the trends we discuss below.

Firstly, by comparing Illustris (red lines) and Illustris-Dark (black lines) in general, we note that the abundances in Illustris are lower relative to Illustris-Dark, consistent with our previous results on the suppression of low-mass subhaloes in haloes of this mass. 
Secondly, we confirm previous DMO results at $z=0$: halo formation redshift and concentration are negatively correlated with subhalo abundances. 
Previous studies on subhalo abundance have often neglected the effect of DM halo shapes on subhalo abundance.  In Illustris-Dark, we find a very weak correlation between halo shape and subhalo abundance: more spherical haloes tend to contain more subhaloes, with such an effect being strongest when using the shape parameter $s$ at $r=0.15 R_{200}$, i.e. relatively close to the halo center. Such a correlation, if it does exist,  can arise from the relationship between halo shape and its orbital characteristics: particles and subhaloes in more spherical haloes have been found in simulations to have a higher proportion of tube orbits to box orbits \citep[][Chua et al. in prep]{Bryan12v422}. Subhaloes in tube orbits tend to have larger pericenter distances, which may reduce the amount of tidal stripping they experience \citep{Wadepuhl11v410}. 

It needs to be noted that all host properties are somewhat correlated with each other, and such correlation could be quantitifed with a Principal Components Analysis (PCA). However, here we want to focus on the dependence of the subhalo abundance with each of them separately, at fixed halo mass. Such an analysis on the correlation of host properties will be examined in future work, and we direct our attention to three main host properties: halo formation time, halo concentration and halo shape.

Since halo formation times are only slightly affected by baryons (see Figure \ref{fig:haloprop}), the trends in the left panels are similar between Illustris and Illustris-Dark. On the other hand, the concentration parameter, though correlated with the formation time, exhibits different trends: Illustris-Dark hosts exhibit strong anti-correlations between current abundances and concentration ($r\sim-0.35$), but this relation is almost absent in Illustris ($r\sim -0.1$ and red solid curve in the top middle panel being almost flat). As already noted in Figure \ref{fig:haloprop} for this mass range, the concentrations of Illustris haloes are offset towards higher values than Illustris-Dark, reflecting  baryonic contraction in the mass density profiles of these haloes.
Finally, the DMO relation between $s$ and $N$ is much less apparent in Illustris.

The results involving halo formation time $z_{1/2}$ are particularly interesting because of the opposite correlations with the current and infall abundances: 
while the current abundance depends inversely on $z_{1/2}$, we find a positive correlation between the infall abundance and $z_{1/2}$.
In other words, for haloes of the same mass at $z=0$, early formers accrete more objects throughout their histories than late formers. This means that late formers accrete fewer, but more massive objects (i.e. more major mergers) than early formers, which has been observed in $N$-body simulations \citep[e.g.][]{McBride09v398}.
On the other hand, the early accreted subhaloes are subjected to a longer duration of tidal and ram pressure stripping that reduce their masses and eventually destroy them \citep{Gao04v355}. 
These two factors counteract each other, and the final abundance at current time is a combination of both increased late-time mergers in late formers and stripping.

We also note that part of the positive correlation between infall abundance and formation time is partially a result of our definition of the normalized accretion mass $\mu_{\rm acc}$, which is normalized using the host mass at $z=0$.
At a fixed current mass, haloes that formed earlier reached larger masses at an earlier epoch and accreted the bulk of their satellites at earlier times with respect to haloes with more recent halo formation times. Given that the masses of the accreted objects are normalized by the mass of the host halo at $z=0$, the bulk of such normalized masses tends to be larger for a halo that formed early (which has therefore grown little since then) compared to a halo that formed late (which has therefore grown significantly in recent times), reducing the normalized masses of satellites accreted throughout its history. We verified using an alternative definition of $\mu_{\rm acc}$ i.e. using both the subhalo and host mass at accretion that this effect accounts for only some of the observed correlation.

The infall abundance remains independent of the concentration regardless of whether baryons were present in the simulations, while current hosts DM shapes are mildly positively correlated with the number of ever accreted subhaloes.
Our results for the concentration parameter are in contrast with those of \cite{Jeeson11v415}, who found from a Principal Component Analysis (PCA) of $N$-body haloes that the DM concentration can be even more fundamental than the halo mass. The lack of correlation between the infall subhalo abundance and the concentration indicates that the concentration might not be as fundamental as \cite{Jeeson11v415} concluded, and reflects the lack of a physical connection between the concentration and infall abundance, with the concentration becoming relevant only after infall. Yet, a correlation at infall might be detected because of some degree of correlation between pairs of host halo properties themselves, as has been observed between halo formation time and concentration and between halo formation time and halo shape%
\footnote{ 
We have also looked into the effect of environments on the SHMF, but did not find any significance dependence in neither the current and infall abundance.
The environmental indicators we investigated included the distance to the closest massive haloes, the environmental measurement $D_{n,f}$ described by \cite{Hass12v419} and the galaxy overdensity $1+\delta$ given in \cite{Illustris}.
This contrasts with results from other simulations possibly due to differing definitions of not only the environment, but also halo mass and subhalo abundance.
For example, \cite{Croft12v425} calculated halo masses using all particles in the FOF group and all subhaloes found within the FOF while \cite{Wang11v413} parameterized the environment by diagonalizing the halo tidal tensor. A more comprehensive study comparing these differences would be required to fully understand how the environment can affect the SHMF.}.

For Milky Way-like haloes, halo formation time seems to be the main driver of the subhalo abundance, and concentration correlates strongly with abundance only for DMO hosts at $z=0$. Yet, it is unclear from our analysis whether, for Milky Way-like haloes, host properties alone -- like the ones studied here, which only depend on the hierarchical growth of structure -- are able to fully justify the amplitude of the scatter in subhalo abundance. \\

\begin{figure}
	\centering 
	\includegraphics[width=0.47\textwidth]{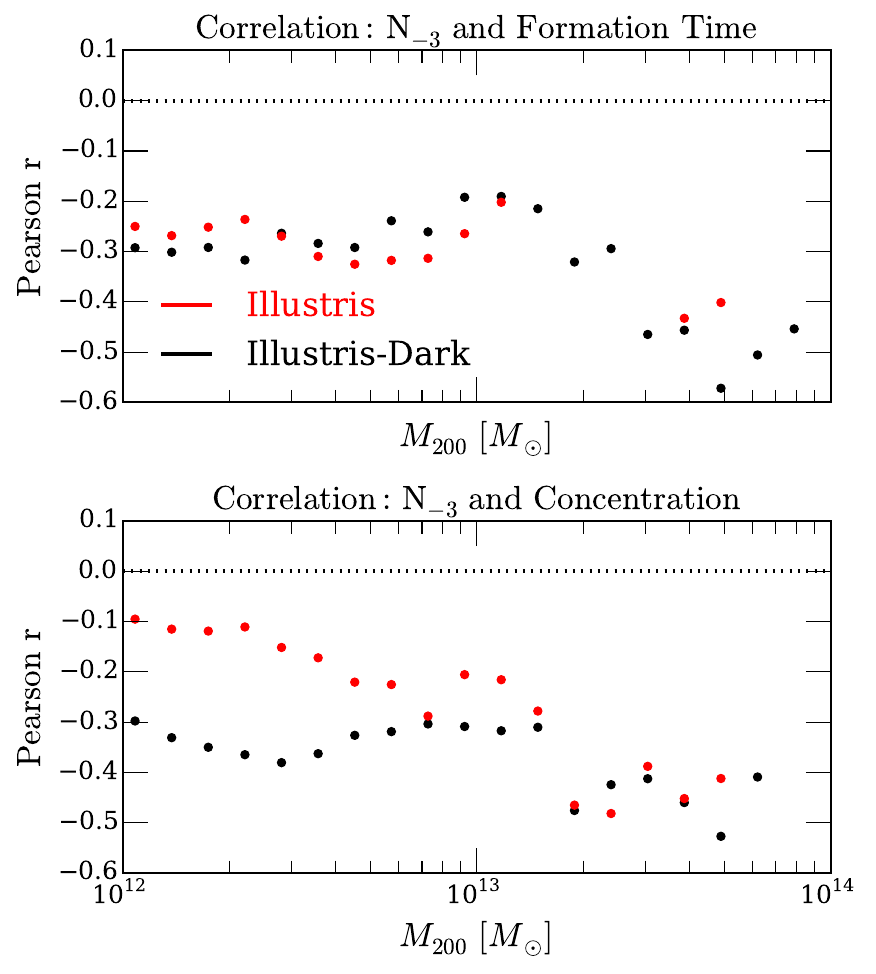}
	\caption{Pearson correlation coefficient between subhalo abundance ($N_{-3}\equiv N_{\rm current}>10^{-3}$) and halo formation time (top) and concentration (bottom) as a function of host halo mass at the current epoch. Red and black curves are for Illustris and Illustris-Dark respectively. Only results which we trust ($p$-value $<$ 0.05) are reported.
	}
	\label{fig:corr_M200}
\end{figure}

We conclude our analysis by extending the characterization of the scatter in  subhalo abundances for different host masses. In Figure \ref{fig:corr_M200}, we show the Pearson correlation coefficients between subhalo abundance and halo formation time (top) and concentration (bottom), respectively, for Illustris (red curves) and Illustris-Dark (black). We use a tophat moving mean and measure the correlation coefficients between the paired quantities in bins of host mass, so that the number of hosts in each bin is not too small. Still, the number of hosts varies from a few hundreds at the low-mass end to a few at the high-mass end, but we do not want to increase the mass bin size in order to avoid mass-dependent effects. Therefore, in Figure \ref{fig:corr_M200}, we only report values of the Pearson coefficients which we trust based on the $p$-value of the correlation: if the correlation is very close to zero and the data points are too few, we do not report any result. In this case, we use a 5\% significance level or equivalently, a threshold of $p$-value $<$ 0.05 for correlations to be considered significant.

Considering subhaloes with $\mu > 10^{-3}$ at $z=0$, we find in the bottom panel of Figure \ref{fig:corr_M200} that red points are almost always closer to 0 than black curves: in Illustris, baryonic physics seems to weaken, slightly, the correlations between subhalo richness and host concentration. On the other hand, the correlation between abundance and halo formation time hold similarly with and without baryonic physics. This demonstrates that the conclusions for Milky Way-like haloes drawn from Figure \ref{fig:propcorr} hold true also at larger host masses (up to about a few $10^{13}M_\odot$). We have checked that similar conclusions hold also for less massive subhaloes, i.e. smaller $\mu$, although the range of host masses for which this can be verified for is very small.

In conclusion, at fixed host masses, among DM halo shape, concentration and halo formation time, we find that halo formation time is the major driver of subhalo abundance in both Illustris and Illustris-Dark; however, we cannot exclude that host physical properties which we have not studied here may have an effect, such as those depending on the gas and stellar content of haloes as well as physical changes within the subhaloes themselves between DMO and FP runs.

\section{Discussion and Conclusions}

We have used the large-scale cosmological hydrodynamical simulation Illustris and its dark-matter only counterpart, Illustris-Dark, to study the abundance of subhaloes, without distinguishing between dark and luminous ones. In particular, we have quantified the average cumulative subhalo mass function (SHMF) and its halo-to-halo variations in Illustris and Illustris-Dark, both at the current epoch and at infall, across a wide range of host ($10^{12}-\text{\rm a few }10^{14} \,M_\odot$) and subhalo total masses ($\gtrsim10^{9} \,M_\odot$). We have compared the findings between the full-physics and the dark-matter only run, validated them against resolution effects, and investigated a selection of possible physical mechanisms responsible for the variations in the average abundances due to baryonic physics and for the diversity of subhalo richness across hosts with similar mass.\\

Our main quantitative results are summarized as follows:

\begin{enumerate}
\item With the Illustris galaxy formation model, the total abundance of subhaloes at $z=0$ is suppressed by baryons in the full-physics run compared to Illustris-Dark. More specifically, baryonic physics tilts the $z=0$ cumulative SHMFs for host masses in the range $10^{12} - 10^{14.5}\msun$, by reducing the abundance of small subhaloes ($\lesssim 10^{10} \msun$) with respect to the DMO case and overall making the function more shallow as a function of subhalo mass.\\

\item We find that baryonic effects are manifest already at infall, i.e. in the abundance of all subhaloes ever accreted by host haloes identified at the current epoch (unevolved SHMF).\\

\item The gravity-only infall SHMF of Illustris-Dark is universal or self-similar, i.e. independent of host halo mass. At $z=0$, even when normalizing by the host mass, low-mass isolated haloes host relatively less subhaloes than the more massive ones, and this effect is more pronounced in Illustris than Illustris-Dark. In fact, the breaking of self-similarity takes place in Illustris already at infall. \\

\item A significant scatter is manifest at fixed host mass, in both Illustris and Illustris-Dark, and both at infall and at the current epoch. For example, in Illustris-Dark at $z=0$ Milky Way-like haloes can host between 3 and 10 subhaloes above $10^{9} \,M_\odot$, within 1-$\sigma$. In general, the subhalo occupation distribution is super-Poissonian for mass ratios $\mu \lesssim 10^{-3}$ and sub-Poissonian for mass ratios $\mu \gtrsim 10^{-2}$, with deviations from Poissonity decreasing from infall to $z=0$. After taking into account variations due to Poisson fluctuations, the normalized intrinsic scatter at fixed host halo masses increases from infall to the current epoch, and from more massive to less massive hosts. Interestingly, the inclusion of baryons does {\it not} entail an increase in the halo-to-halo variation and the intrinsic scatter, at any epoch or mass.\\

\item The non-universality of the average subhalo abundances at infall in Illustris suggests that baryonic effects (which are not scale-free as gravitational collapse) are already in place {\it before} subhaloes accrete onto more massive ones. Indeed, we find that the masses of isolated haloes are different in Illustris and Illustris-Dark both at $z=0$ and at earlier epochs: baryonic processes alter the mass of haloes in Illustris in a non-monotonic manner, generally suppressing the mass with respect to Illustris-Dark except for haloes around $10^{11}\,M_\odot$. The same statement holds for subhaloes at the current epoch.\\

\item  Yet, the suppression of the total average subhalo abundance at current time cannot be explained solely by a systematic modification of subhalo masses, and we observe that more subhaloes in Illustris are totally disrupted and shredded after infall than in Illustris-Dark. This points towards a somewhat enhanced efficiency of Illustris hosts at stripping their satellite haloes, in certain mass ranges.\\

\item We find that not just the mass but also other host properties can be altered by baryonic physics: while Illustris and Illustris-Dark haloes have similar halo formation redshifts, but for a slight shift, baryonic physics in Illustris dramatically alter the concentration--mass relation of isolated haloes at $z = 0$, with objects around $10^{12} \,M_\odot$ and above being more concentrated than their dark-matter only analogs. We speculate that the non-monotonic modification of the concentration--mass relation due to baryonic physics is the main reason for the non-monotonic alteration of the Illustris average SHMF at $z = 0$.\\

\item Finally, {\it at fixed host masses}, among DM halo shape, concentration and halo formation time, we find that halo formation time is the major driver of the abundance halo-to-halo variation in both Illustris and Illustris-Dark, and across (sub)halo masses: haloes that formed earlier host fewer subhaloes than their analogs that formed later with final similar mass.\\

\end{enumerate}

Our findings mean that, for example, a Milky Way-like halo today may host $4.05\pm2.23$ or $6.59\pm2.93$  subhaloes more massive than $10^{9} \,M_\odot$, in a Universe with or without baryons, respectively (within 1-$\sigma$); but a Virgo-like cluster may host today between 85 and 95 subhaloes exceeding  $10^{10} \,M_\odot$ in Illustris instead of about $98-110$ similar satellites in Illustris-Dark (within the 1-$\sigma$ scatter). Moreover, because of the different levels of DM halo concentration in Illustris vs. Illustris-Dark, a Milky Way mass object in Illustris may be more resistant to (tidal) stripping if it is a subhalo of a larger host, while, if a host halo, it may induce an enhanced (tidal) stripping and mass loss into its satellites, when compared to Illustris-Dark. At the same time, also in full baryonic simulations like Illustris, a Milky Way-like host that formed more recently will host a larger number of subhaloes with respect to a similarly massive and concentrated halo which formed earlier because, in the former case, the bulk of subhaloes accreted more recently and hence had a shorter time to undergo subhalo-host interaction processes, such as tidal stripping.

We postpone to future efforts the task of quantifying the impact and mass-dependence of other physical processes which we have only mentioned in this paper. These processes may depend on the stellar and gaseous contents of (sub)haloes, other baryons-induced changes to the inner structures of subhaloes and ram-pressure stripping. These may act in addition to the ones we have identified here -- mainly DM halo concentration and halo formation time --  and may bring a more detailed understanding to both the baryonic alteration of the average subhalo abundances across cosmic times as well as their halo-to-halo variations. 

Our results have implications for the gamma ray flux expected from supersymmetric dark matter annihilation in Milky-Way sized haloes. $N$-body simulations have previously predicted the total DM annihilation flux to be dominated by emission from substructures \citep[e.g.][]{Diemand07v667,Springel08v456}. Reduced total substructure abundance, as we find in Illustris, could result in a lower contribution to the annihilation flux from substructure emission and larger contribution from smooth emission from the main halo instead. However, we do not expect the predicted detectability of DM annihilation signals from large subhaloes to be significantly affected by baryons since their abundances are not decreased going from Illustris to Illustris-Dark.

\section*{Acknowledgements}
We thank Volker Springel, Fangzhou Jiang and the referee for their constructive comments on the paper. AP thanks Joel Primack and Peter Behroozi for early access to halo catalogs of the Bolshoi simulations used to verify literature $N$-body results in Section \ref{sec:scatter}.
The simulations analyzed in this paper were run on the Harvard Odyssey and CfA/ITC clusters, the Ranger and Stampede supercomputers at the Texas Advanced Computing Center as part of XSEDE, the Kraken supercomputer at Oak Ridge National Laboratory as part of XSEDE, the CURIE supercomputer at CEA/France as part of PRACE project RA0844, and the SuperMUC computer at the Leibniz Computing Centre, Germany, as part of project pr85je. AP acknowledges support from the HST grant HST-AR-13897.

\appendix
\section{Resolution Convergence and additional results}
\label{sec:appendix}

 \begin{figure*} 
 	\centering 
 	\includegraphics[width=\textwidth]{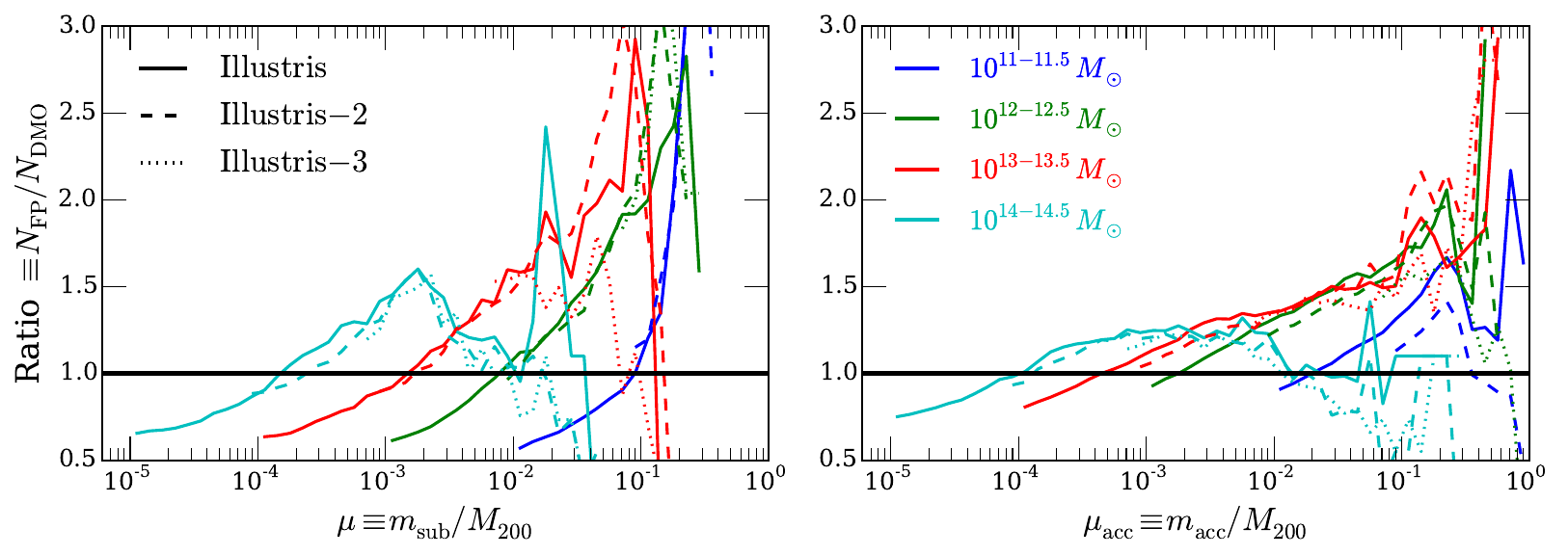}
 	\caption{Resolution convergence of the effects of baryons on the subhalo demographics. Solid curves: ratio of the mean SHMF in Illustris to that of Illustris-Dark  as a function of the normalized subhalo mass, identical to the bottom panels of Figure \ref{fig:shmf}. Results for the lower resolution Illustris-Dark2 and Illustris-(Dark)3 are shown as dashed and dotted curves, respectively.
 	Left and right panels show the ratios of the current SHMF and infall SHMF respectively. 
 	}
 	\label{fig:shmfres}
 \end{figure*}

We show that our results for the subhalo mass function are converged with resolution with the subhalo mass cut we chose in this work. 
The effect of resolution on baryonic physics is shown in  Figure \ref{fig:shmfres}, plotting the ratio of the SHMF of Illustris to Illustris-Dark (similar to the lower panels of Figure \ref{fig:shmf}).
Solid, dashed and dotted lines correspond to the highest resolution Illustris, and the lower resolution Illustris-2 and Illustris-3, respectively.
The mass resolution elements are larger in the lower resolution runs, leading to minimum subhalo masses of $8\times10^9 \msun$ and $6.4\times10^{10} \msun$ in Illustris-2 and Illustris-3 respectively (including their dissipationless dark counterparts).
This restricts the range of $\mu$ to larger values.
Our results show good agreement between all three resolution runs in each host mass bin.

\section{Fitting for the Subhalo Mass Function}
\label{sec:shmffit}

 \begin{figure*} 
	\centering 
	\includegraphics[width=0.49\textwidth]{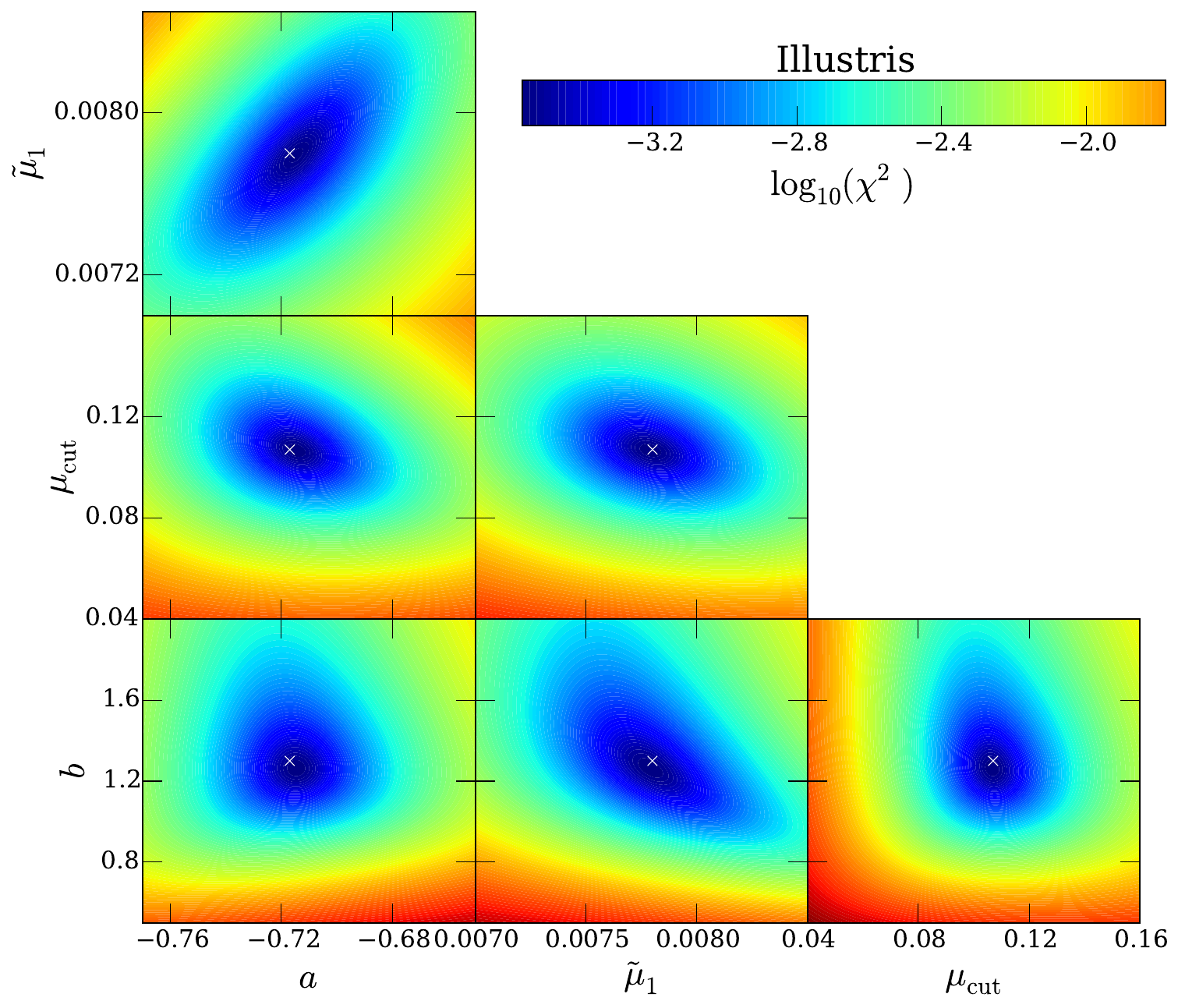}
	\includegraphics[width=0.49\textwidth]{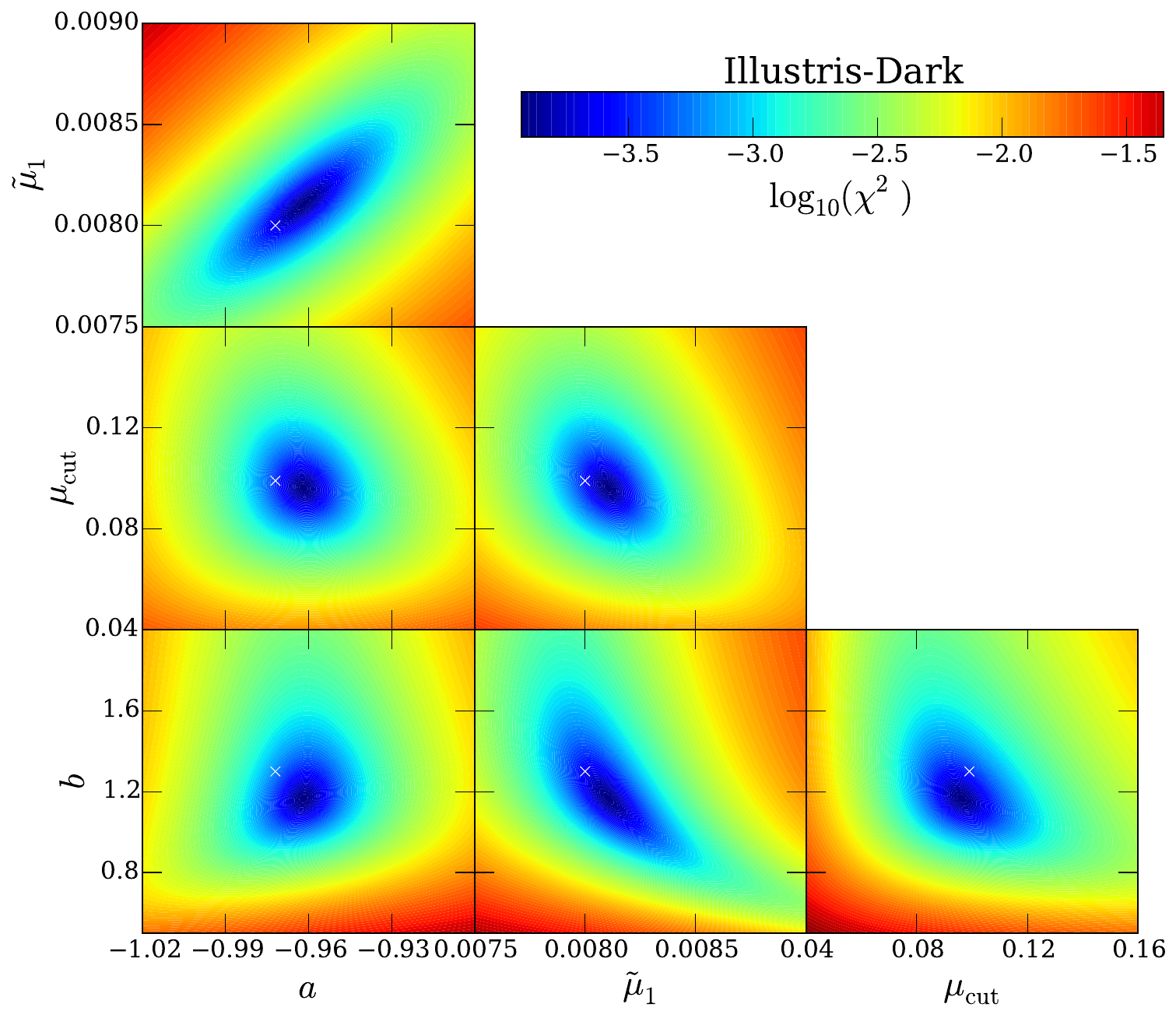}
	\caption{
	Degeneracy of subhalo mass function fitting parameters (equation \ref{eq:cdf}) at current epoch for $10^{12} \msun$ hosts in Illustris (left) and Illustris-Dark (right). Colour indicates the $\chi^2$ as defined by equation \ref{eq:chi2}. The white crosses mark the best fit parameters obtained from fixing $b=1.3$. We find that the choice of fixing $b=1.3$ for the current epoch is a reasonable one. At the same time, the degeneracy between parameters is generally stronger in Illustris-Dark than Illustris.
	}
	\label{fig:shmffits}
\end{figure*}

In section \ref{sec:shmf}, we provided fitting parameters for the cumulative SHMFs to equation \ref{eq:cdf}, with the parameter $b$ fixed at $b=1.3$ for current time and $b=0.8$ at infall. We show in Figure \ref{fig:shmffits} the 2D posterior probability plots for haloes of mass $10^{12-12.5} \msun$, allowing all four parameters in equation \ref{eq:cdf} to be varied. The left and right panels show the results from Illustris and Illustris-Dark respectively, with colour representing $\chi^2$ values (as calculated using equation \ref{eq:chi2}), blue being the minimum $\chi^2$. The white crosses mark the best fit parameters obtained from fixing $b=1.3$.

From Figure \ref{fig:shmffits}, we note that the degeneracy between parameters is stronger in Illustris-Dark compared to Illustris. For $10^{12} \msun$ haloes in Illustris, the inferred parameters obtained from fixing $b=1.3$ lie near the minimum $\chi^2$, confirming that our choice of fixing $b=1.3$ is a good one. For Illustris-Dark, the agreement is poorer, but the small $\chi^2$ values still indicate a reasonable fit. We confirm independently from a plot of the SHMF that the fits reproduce the shape of SHMF curves in both Illustris and Illustris-Dark reasonably well.

{\bibliographystyle{mnras}
\bibliography{references}
}

\end{document}